\documentclass[a4paper,11pt]{article}
\pdfoutput=1

\usepackage{jcappub}

\usepackage{graphicx}
\usepackage{multirow}
\usepackage{slashed}
\usepackage{gensymb}
\usepackage[table]{xcolor}
\usepackage[T1]{fontenc}
\usepackage{subfig}
\usepackage{caption}
\usepackage{graphicx}
\usepackage{amsmath}
\usepackage{makecell}
\usepackage{amssymb}
\usepackage{latexsym}
\usepackage[percent]{overpic}

\title{\boldmath Flavorful Interactions of AGN Neutrinos  with Dark Matter Spike}

\author[]{S. Abbaslu,}
\author[1]{Y. Farzan\note{Corresponding author}, }
\author[]{ and A. Maleki}

\affiliation[]{School of physics, Institute for Research in Fundamental Sciences (IPM) \\  P.O.Box 19395-5531, Tehran, Iran}

\emailAdd{s-abbaslu@ipm.ir}
\emailAdd{yasaman@theory.ipm.ac.ir}
\emailAdd{ali.malekiphys@gmail.com}

\abstract{IceCube-Gen2 is going to make the dream of precise flavor ratio measurement for high energy cosmic neutrinos a reality. Motivated by this prospect, we build a model for the interaction of neutrinos with the dark matter and study the impact of the neutrino interaction with the dark matter spike around active galactic nuclei on the  neutrino flavor ratio measurement.  We show that, under certain circumstances,  the flavor measurement by IceCube-Gen2 may discriminate between this model and the standard expectation, $(\nu_e^\oplus:\nu_\mu^\oplus:\nu_\tau^\oplus)\simeq (1/3:1/3:1/3)$, as well as the prediction for a damped muon source.  We discuss how we can derive information about the spike as well as about the characteristics of the dark matter particles composing it by combining the flavor ratio measurements with the results of the terrestrial experiments determining the neutrino mass ordering and a potential deviation from the standard model predictions in the measurements of the tau decay modes.}

\begin{document}
\maketitle
\flushbottom

\section{Introduction}

As shown by \cite{Gondolo:1999ef}, during the formation of supermassive  black holes in the center  of the galactic dark matter halos, the distribution of dark matter in the proximity of the black hole
is severely affected, leading to a  profile in the vicinity of the black hole denser than the  galactic dark matter  halo, known as the ``spike."
Ref. \cite{Sharma:2025ynw} has found observational  hints in favor of the dark matter spikes around certain supermassive black holes.
In recent years, rich literature has been developed both on the formation  and on the profile of the spike. Possible interactions between dark matter particles inside the spike and the proton and/or electron jets or the high energy neutrino flux coming out of Active Galactic Nuclei (AGN) have also been scrutinized.
Moreover, the impacts of dark matter pair annihilation both on the profile of the spike \cite{Vasiliev:2007vh,Shapiro:2016ypb} and on the indirect dark matter search experiments \cite{Chattopadhyay:2026kbm} have been widely studied. Refs. \cite{DeMarchi:2025xag,DeMarchi:2025uoo} suggest the deep inelastic scattering of the high energy cosmic ray coming out of AGN off the dark matter particles composing the spike as the origin of the high energy neutrino flux observed by the IceCube and KM3NeT neutrino telescopes. Ref. \cite{Fujiwara:2024qos} points out that the neutrino dark matter interaction can change the diffusion time of the neutrinos out of the transient sources.

In Refs. \cite{Cline:2022qld,Cline:2023tkp,Zapata:2025huq,Dixit:2026zkv,Mondol:2025uuw}, bounds are set on the cross section of neutrino scattering off dark matter based on the consideration that for large scattering  rates, the neutrino flux from the AGN sources would be attenuated against the observation made by IceCube.  In this paper, we are interested in relatively large couplings but as we shall see, because of the flavor structure of the coupling that we consider, a severe attenuation of the flux can be avoided.   We explore the implication of the interactions between the AGN neutrino flux  and the dark matter particles composing the spike for the flavor ratios at the Earth.

Let us denote the dark matter particles by $\psi$. While Refs. \cite{Cline:2022qld,Cline:2023tkp,Dixit:2026zkv} focus on the elastic scattering $\nu+\psi\to \nu+\psi$, we discuss inelastic scattering $\nu +\psi \to N+\psi$ where $N$ is a Heavy Neutral Lepton (HNL).
We show that the subsequent (chain) decay of HNL to $\nu \bar{\nu}\nu$  can lead to interesting phenomenological observations.
We take $\psi$ to be  a fermion charged under a new $U(1)$ gauge symmetry with a light gauge  boson, $V$.   We also  couple $V$ to $\nu$ and to $N$, off-diagonally;  such that the $\nu+\psi  \to N +\psi$ scattering can take place via a $t$-channel $V$ exchange.  Since the mediator is taken to be a vector boson (rather than ${\it e.g.}$ a scalar), at large $\nu$ energies (even much larger than $m_V^2/2m_\psi$), the scattering cross section remains large (given by $m_V^{-2}$ rather than by $(2 E_\nu m_\psi)^{-1}$), with $d\sigma /dE_\nu$ peaked around $E_N \simeq E_\nu$, corresponding to an energy momentum transfer of $O(m_V)$. 

We demonstrate that with symmetric dark matter ({\it i.e.,} with equal densities for $\psi$ and $\bar{\psi}$ in the spike),  the same coupling that leads to the scattering will also lead to the annihilation of dark matter pair in the inner part of the spike. We argue that  although invoking a co-annihilation scenario can reduce the annihilation cross section but despite even simultaneous $p$-wave and loop suppressions, the rate of the dark matter pair annihilation to $\nu\bar{\nu}$ will be enough to reduce the column densities in the spike to an extent that renders the neutrino scattering inefficient. We therefore focus on the asymmetric dark matter scenario \cite{Nussinov:1985xr,Barr:1990ca,Barr:1991qn,Kaplan:1991ah,Davoudiasl:2012uw,Zurek:2013wia, Dev:2026ddq} in which the neutrinos at the production point at the spike obtain a large  dark matter induced matter potential, drastically changing their flavor evolution even without hard scattering.  We assume the new physics preserves lepton flavor but we take the couplings of different neutrino flavors non-universal.  We show that the neutrino evolution in the spike will be adiabatic, implying that without hard scattering, a pure flavor state will exit the spike as a pure mass eigenstate. If  only $\nu_\tau$ couples to the gauge boson, there will be no hard scattering and therefore no attenuation of the flux or distortion of the energy spectrum but as we shall see, the flavor ratios at the Earth or the ratio of the $\nu_\mu$ flux to that of $\bar{\nu}_\mu$ can be discerned from the standard prediction by IceCube-Gen2. When we also turn on the coupling of  $\nu_e$ or $\nu_\mu$, hard scattering and therefore attenuation of the flux can take place but as we shall see, since at most one flavor undergoes scattering, the attenuation will be mild and can be hidden in the uncertainties of the AGN models predicting the  luminosity of the neutrino flux 
\cite{Inoue:2019yfs,Murase:2019vdl,Kheirandish:2021wkm,Murase:2022dog,Eichmann:2022lxh,Inoue:2022yak,Yoast-Hull:2013qfa,Fiorillo:2023dts,Fiorillo:2024akm}.  We discuss  the predictions for  the flavor ratios on the Earth which turn out  to strongly
depend on the neutrino mass ordering. In the majority of the cases, the prediction deviates from the standard expectation to an extent that can  be resolved by IceCube-Gen2. We discuss how by combining the information from the upcoming terrestrial neutrino and charged lepton experiments  and the flavor ratio measurement by IceCube-Gen2, mysteries of the spikes and dark matter particles building them can be unraveled.

The paper is organized as follows. In sect. \ref{spike}, we describe the properties of 
the typical spikes around AGN neutrino sources. We then present the outlines of  the scenario for the neutrino interaction with the dark matter in the spike, introducing the cascade equation governing  the evolution of the neutrino flux propagating in the spike for the single flavor case. In sect. \ref{model}, we present the dark matter model and discuss the present bounds on its parameters and then highlight the ranges of parameters that can give rise to non-trivial observable effects on the  neutrino flux traversing the spike. In sect.~\ref{ice}, we review the results of IceCube for the flavor ratio measurement as well as the IceCube-Gen2 forecast for flavor probing. In sect. \ref{medium}, we study the flavor evolution in the presence of dark matter induced   matter potential for the neutrinos. We then show the predictions for the flavor ratios taking into account both the effects of hard and forward scattering off the  dark matter in the spike.  We discuss  the implications of different combinations of IceCube-Gen2 observations and the upcoming measurements (and hopefully discoveries) by terrestrial experiments.  We summarize the results in sect.~\ref{con} and further discuss the roadmap for future studies in this direction. In appendix~\ref{depletion}, we show that even in the case that all neutrinos go through hard scattering at the spike, they cannot destroy the spike. In appendix~\ref{cascade}, we derive the energy spectrum of the final neutrinos from the chain decay, $N\to \nu V$ followed by $V\to \nu \bar{\nu}$. We then generalize the cascade equation to include the flavor structure of the interactions.

\section{Neutrino flux evolution through a spike  \label{spike}}

Let us consider an AGN with a central black hole of mass, $M_{BH}$ and  Schwarzschild radius, $R_S=2 G_N M_{BH}$.
For $r\gg 4R_S$, the density profile of a spike with radius $R_{SP}$ can be parameterized as 
\begin{equation}
	\rho_{spike}(r)= \rho_{halo} (R_{SP})\left(\frac{r}{R_{SP}}\right)^{-\gamma_{SP}}  \ \ \ {\rm for } \ \ r<R_{SP} \ . \label{spike-profile}
\end{equation}
The value of $\gamma_{SP}$ depends on various conditions including (non-)adiabacity of the black hole formation \cite{Ullio:2001fb,Balaji:2023hmy}, stellar  heating \cite{Gnedin:2003rj,Shapiro:2022prq,Bertone:2005hw}, the density profile of the halo and etc. However, its value is known to be larger than 1 which means as we approach the central black hole, the dark matter density increases even faster than the halo profile. 
For $r>R_{SP}$, the dark matter distribution is described by the dark matter halo profile. By imposing continuity at the intersection point where the inner spike density matches that of the dark halo, the spike radius can be determined. See Refs \cite{Dixit:2026zkv,Cline:2023tkp} for more details. 
If the dark matter particles annihilate with each other,  at small $r$ where the annihilation rate approaches the inverse of the spike age \cite{Vasiliev:2007vh,Shapiro:2016ypb},  the density  becomes
saturated. Within an asymmetric dark matter scenario, the dark matter particles inside the spike cannot annihilate so the spike density will  continue  to increase down to $\sim 4 R_S$ as described in Eq. (\ref{spike-profile}).

The distance of the neutrino production region from the black hole is denoted by $R_{em}$ which can be as low as $30 R_s$
\cite{Murase:2019vdl}.  
In the literature studying neutrino dark matter interaction \cite{Cline:2022qld,Cline:2023tkp}, to compute the column density through which neutrinos propagate in the spike, the lower limit of integration should be taken equal to $R_{em}$:
\begin{equation}
\Sigma\equiv \int_{R_{em}} ^{R_{SP}} \rho_{spike}(r) dr\ . \label{column}
\end{equation}

\begin{table*}[ht!]
	\caption{
		Black hole and spike parameters for NGC-1068 \cite{IceCube:2022der}, NGC4151 \cite{IceCube:2024ayt}, TXS+0506+056 \cite{IceCube:2018dnn} and PKS 1424+240 \cite{IceCube:2022der}.
		$\Sigma_{7/3}$ and $\Sigma_{3/2}$ are  the column densities in $\rm{GeV/cm^2}$ computed using Eq. (\ref{column}) for two values of $\gamma_{SP}$, $7/3,3/2$. The corresponding spike densities, $\rho_{7/3}$ and $\rho_{3/2}$ in $\mathrm{GeV/cm^3}$, are evaluated at $r=R_{\rm em}$.}
	\label{table:parameters}
	\setlength{\tabcolsep}{11pt}
	\rowcolors{1}{gray!10}{cyan!10}
	\begin{tabular}{lcccc}
		\hline\hline
		Parameters & NGC 1068 & NGC 4151 & TXS 0506+056 & PKS 1424+240 \\
		\hline\hline
		$M_{\rm BH}$ ($M_{\odot}$) 
		& $1.0\times10^{7}$
		& $2.0\times10^{7}$
		& $3.09\times10^{8}$
		& $1.0\times10^{9}$ \\
		\hline
		$R_S$ (pc)
		& $9.5\times10^{-7}$
		& $1.9\times10^{-6}$
		& $2.95\times10^{-5}$
		& $9.5\times10^{-5}$ \\
		\hline
		$R_{\rm em}$ (pc)
		& $2.85\times10^{-5}$
		& $5.7\times10^{-5}$
		& $6.25\times 10^{-2}$
		& $6.36\times 10^{-2}$ \\
		\hline
		$R_{SP}^{(\gamma_{SP}=7/3)}$ (pc)
		& $674.1$
		& $295.81$
		& $159.9$
		& $503.18$ \\
		\hline
		$R_{SP}^{(\gamma_{SP}=3/2)}$ (pc)
		& $1236.66$
		& $543.19$
		& $ 6027.83$
		& $8416.85$ \\
		\hline
		$\Sigma_{7/3}$
		& $5.13\times10^{31}$
		& $8.62\times10^{30}$
		& $1.16\times10^{27}$
		& $7.8\times10^{27}$ \\

		\hline
		$\Sigma_{3/2}$
		& $2.27\times10^{28}$
		& $3.35\times10^{29}$
		& $9.57\times10^{24}$
		& $8.1\times10^{25}$ \\

		\hline
		$\rho_{7/3}(R_{em})$
		& $6.24\times10^{17}$
		& $5.25\times10^{16}$
		& $8.3\times10^{9}$
		& $5.4\times10^{10}$ \\

		\hline
		$\rho_{3/2}(R_{em})$
		& $3.2\times10^{14}$
		& $2.3\times 10^{15}$
		& $7.6\times 10^{7}$
		& $6.3\times 10^{8}$ \\
		
		\hline\hline
	\end{tabular}
\end{table*}

Table  \ref{table:parameters} shows the values of $\rho_{spike}(R_{em})$ as well as  those of $\Sigma$ for the four observed point sources of high energy neutrinos, NGC-1068 \cite{IceCube:2022der}, NGC4151 \cite{IceCube:2024ayt}, TXS+0506+056 \cite{IceCube:2018dnn} and PKS 1424+240 \cite{IceCube:2022der}, taking   $\gamma_{SP}=7/3, 3/2$ which are  theoretically motivated indexes. The values of  $M_{BH}$, $R_S$, $R_{SP}$ and $R_{em}$ are also shown in this table \cite{Greenhill:1996te,Bentz:2022NGC4151,Padovani:2019xcv,Cerruti:2017mnz}.

Throughout our analysis, we shall take NGC 1068 to be a typical AGN contributing to the high energy neutrino flux.  Indeed, as we shall discuss, collecting the events originated only from this powerful and steady source of high energy neutrinos  (NGC 1068) can provide rich information on the new physics suggested in the present paper. The value of the optical depth is strongly sensitive to the lower limit of the integration  as well as to $\gamma_{SP}$ which  depends on the specific properties of the AGN (or the halo hosting it)
and may differ for different AGNs. For $\Sigma  \sigma /m_{\psi} \stackrel{>}{\sim} 1$ with $\sigma$ being the neutrino scattering cross section off dark matter, a significant fraction of neutrinos undergo scattering before leaving the spike. This may attenuate the neutrino flux \cite{Cline:2022qld,Cline:2023tkp,Dixit:2026zkv,Dev:2025tdv}. The amount of the attenuation however depends not only on $\Sigma \sigma/m_{\psi}$ but also on the details of the model for such scattering \cite{Farzan:2026wek}. As we shall see in our model, scattering replaces each neutrino with three lower energy neutrinos.   It is straightforward to show that the column density encountered by the neutrinos after exiting the spike down to the Earth is negligible in comparison to that of the spike; {\it i.e.,} $\int \rho dr$ along the path in the halo and between two galaxies is smaller than this integration inside the spike.

In our model, the couplings relevant for neutrino scattering off dark matter and its regeneration are the following
\begin{equation}
V_\mu \left( g_\psi \bar{\psi} \gamma^\mu \psi+g_{\nu N}\bar{\nu}_L \gamma^\mu N_L +g_{\nu \nu} \bar{\nu}_L \gamma^\mu \nu_L\right)+H.c. \label{Lag-eff}
\end{equation}
Since we have taken asymmetric dark matter scenario, the $g_{\nu\nu}$ coupling along with $g_\psi$ gives rise to  a dark matter  induced matter potential for the neutrino, similarly to the  standard medium effect: $(g_\psi g_{\nu \nu}/m_V^2)(\rho_{spike}/m_\psi)$. If the dark matter is symmetric with equal number densities for $\psi$ and $\bar{\psi}$, the contributions from $\psi$ and $\bar{\psi}$ to the  dark matter  induced matter potential would cancel each other. As a result, the evolution of the flavor ratios of the flux would be completely different. However, as we shall demonstrate, with symmetric model ($n_\psi=n_{\bar{\psi}}$), dark matter pair annihilation and therefore the destruction of the spike are inevitable.
In the next sections, we will discuss the underlying model as well as the flavor structure of the couplings. For the time being, let us forget about  the neutrino mixing and the flavor issues and discuss  the impact of the  scattering  on the attenuation for a single flavor neutrino flux .
We are interested in the regime $E_\nu \gg m_V, m_{\psi}, m_N$. 
In this regime, up to  negligible corrections of $\mathcal{O}[(m_V^2, m_N^2,m_\psi^2)/E_\nu^2] <10^{-8}$ the total cross section and the diffeential cross section can be written as 
\begin{equation} \label{eq:scatterX}
\sigma\simeq\frac{g_{\nu N}^2g_{\psi}^2}{4\pi m_V^2}\left(1-\frac{m_V^2}{s}\ln\left[1+\frac{s}{m_V^2}\right]\right) ,  
\end{equation}
and
\begin{equation}\label{differential}
\begin{split}
\frac{d\sigma}{dE_N}&\simeq \frac{g_{\nu N}^2g_{\psi}^2}{4\pi  }\frac{m_{\psi}}{(2m_{\psi}(E_N-E_{\nu})-m_V^2)^2}\left(\frac{E_{\nu}^2 +E_{N}^2}{E_{\nu}^2}\right)\ .
\end{split}
\end{equation}
Notice that $2m_\psi(E_N-E_\nu)$ appearing in the denominator is exactly equal to the Mandelstam variable,  $t$.  In other words, the equation  $t=2m_\psi(E_N-E_\nu)$ is exact  including the effects of nonzero $m_N$ and $m_\psi$. It is straightforward to show that the minimum of $|t|$ is $\mathcal{O}[m_N^2 {\rm Max}(m_\psi^2,m_N^2)/E_\nu^2]\ll m_V^2$.
Notice that we have taken $\psi$ to be a full four component Dirac spinor with a vector coupling to $V_{\mu}$. 
The scattering cross sections for neutrinos and antineutrinos in our model are equal. From now on, wherever  we discuss the neutrino flux, we implicitly  mean the  combined flux of neutrinos and antineutrinos as their evolution is governed by the same scattering cross section. Had we taken a chiral coupling for $\psi$, the neutrino and antineutrino scattering cross sections could be different, inducing unnecessary complication for the analysis of interest in the present study but as we shall elaborate in sect.~\ref{model}, it could help in the direction of model building minimalism.

Notice that  because we have taken a vector mediator, even for $E_\nu \to \infty$, the cross section does not vanish and is still given by $1/m_V^2$ rather than by $1/s\sim 1/(2 m_{\psi} E_\nu)$. This is because even for large $E_\nu$, for most of the scatterings the energy momentum transfer is small and of order of $m_V$. From Eq. (\ref{differential}), we also observe that for the most of the scatterings, $(E_N-E_\nu)\sim m_V^2/m_\psi$.  Thus, for $m_V,m_\psi, m_N\sim {\rm few}\times 100$~MeV, $N$ will inherit almost all the energy of the initial neutrino, leaving only a kinetic energy of $m_V^2/m_\psi \sim m_\psi$ for the final $\psi$. At $m_\psi=0.1$~GeV, $m_N=0.3$~GeV and $m_V=0.12$~GeV corresponding to the ${\rm B}_1$ and ${\rm B}_2$ benchmarks ($m_V=0.06$~GeV corresponding to the ${\rm B}_3$ and ${\rm B}_4$ benchmarks), for high energy neutrinos with  $E_\nu>1.5$~TeV$$ \frac{E_\nu-\langle E_N\rangle}{E_\nu}=1-\frac{1}{E_\nu}\frac{\int E_N\frac{d\sigma}{dE_N}dE_N}{\int \frac{d\sigma}{dE_N}dE_N} < 4\times 10^{-4}  \ \ (10^{-4})$$  so we can safely take $E_N=E_\nu$. The average scattering angle in the center of mass frame is $\mathcal{O}(m_V/\sqrt{E_\nu m_\psi})$ which in the lab frame will be  $\mathcal{O}(m_V/E_\nu)\ll 0.1^\circ$ which is much smaller than the typical opening angle of the AGN jets \cite{Pushkarev:2017fbk}. Thus, $N$ will not deviate away from the direction of the jet. Although the final $\psi$ inherits only a small fraction of the initial $\nu$ energy, it can still be highly boosted with a kinetic energy of order of 100 MeV and velocity higher than the escape velocity from the spike. We shall discuss in appendix \ref{depletion} whether this process can deplete the spike or not. If $\psi$ couples to the matter field, it can leave a sizable recoil energy at the scattering in the direct detection experiments \cite{Barillier:2025xct}. In our model, $\psi$ does not however couple to the matter field. 

 Via the $V$ mediator, $N_L$ can immediately decay as  $ N_L \to V^* \nu_\tau \to \nu_\tau \bar{\nu}_\tau \nu_\tau$.  Since $N$ is highly boosted, the average angle that the the momenta of the final neutrinos make with the direction of $N$ is $\mathcal{O}(m_N/E_\nu)\ll 0.1^\circ$ so the final neutrinos will remain within the direction of the initial jet which has a much larger opening angle relative to $\mathcal{O}(m_N/E_\nu)$ \cite{Pushkarev:2017fbk}.

Let us now discuss the attenuation of the neutrino flux. Thanks to the scattering, each neutrino  (anti-neutrino) is replaced by  two (one) neutrinos plus one (two) anti-neutrino so the flux evolution can be described by the following differential equation
\begin{equation} \label{Phi-evolution}
\frac{m_\psi}{\rho_{spike}}\frac{d\Phi_\nu(E_\nu)}{dr}=-\sigma \Phi_\nu(E_\nu)+ \int_{E_\nu}^\infty dE_\nu^\prime \sigma \left(\sum_{i=1}^2\frac{dN_i(E_\nu,E_\nu^\prime)}{d E_\nu} \Phi_\nu(E_\nu^\prime) +\frac{dN_3(E_\nu,E_\nu^\prime)}{d E_\nu} \Phi_{\bar\nu}(E_\nu^\prime)\right)
\end{equation}
where $\Phi_\nu$ and $\Phi_{\bar{\nu}}$ are respectively  the  fluxes of neutrinos and anti-neutrinos  multiplied by $r^2$.  The first term in the right-handed side of Eq. (\ref{Phi-evolution}) accounts for the attenuation due to scattering. The second term accounts for the $\nu$ regeneration due to the $N_L$ decay.   While
$d N_{1,2}(E_\nu , E_\nu^\prime)/dE_\nu$ determine the shapes of the energy spectra of the final neutrinos created by an initial $\nu$ with an energy of $E_\nu^\prime$, $d N_{3}(E_\nu , E_\nu^\prime)/dE_\nu$ shows the spectrum of the final $\bar\nu$. Since the cross sections of $\nu $ and $\bar{\nu}$ as well as the spectra of the decay products of $N$ and  $\bar{N}$ are identical, the evolution of $\Phi_{\bar{\nu}}(E_\nu)$ is also given by Eq. (\ref{Phi-evolution}), replacing $\Phi_\nu \leftrightarrow \Phi_{\bar{\nu}}$.
Depending on whether $N_L$ is heavier or lighter than $V$, the spectrum of final neutrinos will be different. In appendix \ref{cascade}, we have derived $dN_i/d E_\nu$ for the case that $N$ decays first to $\nu$ and an on-shell $V$ and then $V\to \nu\bar{\nu}$. This is the case for our benchmarks at which $m_V<m_{N}$.
As discussed before, for $E_\nu \gg m_V^2/m_\psi\sim 100~{\rm MeV}$, $\sigma$ approaches a constant value in energy: $\sigma=\sigma(E_\nu)$ so it can be factored out of the integration in the generative term in Eq. (\ref{Phi-evolution}). Moreover, as discussed before in this regime, $N$ inherits almost all the energy of the initial $\nu$: $ E_{\nu}^\prime-E_N=\mathcal{O}(m_V^2/(2m_\psi))\ll E_N$.  In appendix  \ref{cascade}, we have shown that $\int_{E_\nu} (dN/dE_\nu )E_N^{-\gamma_\nu} dE_N\propto E_\nu^{-\gamma_\nu}$.
Thus,
taking a power law behavior for $\Phi_\nu$  and   $\Phi_{\bar\nu}$ at the production,
\begin{equation} \Phi_\nu|_{initial}=A^0_{{\nu}} E_\nu^{-\gamma_\nu}  \ \  {\rm and} \ \  \Phi_{\bar\nu}|_{initial}=A^0_{\bar\nu} E_\nu^{-\gamma_\nu}  \\,\label{initials} \end{equation}
the power law behavior will be maintained across the spike:
\begin{equation}
\frac{m_\psi}{\rho_{spike}} \frac{d\Phi_\nu(E_\nu)}{dr}=-\sigma \Phi_\nu+b\sigma \Phi_\nu + \bar{b}\sigma \Phi_{\bar\nu} \  \ {\rm and } \ \ 
\frac{m_\psi}{\rho_{spike}} \frac{d\Phi_{\bar\nu}(E_\nu)}{dr}=-\sigma \Phi_{\bar\nu}+{b}\sigma \Phi_{\bar\nu} +\bar{b}\sigma \Phi_\nu \label{maintain}
\end{equation}
in which
\begin{equation} 
\label{b} 
b=\mathcal{B}+ 
\frac{1}{\gamma_\nu^2}\left(\frac{m_N^2}{m_N^2-m_V^2}\right)\left( 1-\left(\frac{m_V^2}{m_N^2}\right)^{\gamma_\nu}\right) \   {\rm and} \ \bar{b}=\frac{1}{\gamma_\nu^2}\left(\frac{m_N^2}{m_N^2-m_V^2}\right)\left( 1-\left(\frac{m_V^2}{m_N^2}\right)^{\gamma_\nu}\right)
\end{equation}
where $\mathcal{B}$ is defined in Eq. (\ref{matB}). 
For the derivation of the formula  for $b$ and $\bar{b}$, see appendix \ref{cascade}. 
This simple differential equation can be very easily solved.  We shall solve it taking into account the non-trivial flavor structure of the coupling

Another concern is that if the intensity of the flux is high enough, the neutrino scattering may deplete dark matter from the spike. In appendix \ref{depletion}, we show that the luminosity of the flux  is not enough for such depletion.


\section{The model \label{model}}

In this section, we build a dark matter model underlying the interaction in Eq. (\ref{Lag-eff}). We then explore the parameter space that   satisfy all the existing astrophysical and terrestrial bounds and at the same time, lead  to non-trivial effects on the neutrino flux passing through the dark matter spike around AGNs.

As discussed before, $V_\mu$ is the gauge boson of a new $U_{NEW}(1)$ gauge theory. The $U_{NEW}(1)$ charges of the new particles required for this model are displayed in tab \ref{tab:charges2}.

\begin{table}[htb!]
\setlength{\tabcolsep}{7pt} 
\renewcommand{\arraystretch}{1.0} 
\caption{New particles and their relevant quantum numbers under the $U_{NEW}(1)$ gauge symmetry group. All these new particles are singlets under the standard model  (SM) gauge group  and the SM particles are singlets under $U_{NEW}(1)  $. For the case that instead of $\nu_e$,  $\nu_\mu$  mixes with HNL, we may change the name of $N_e$ to $N_\mu$.} 
\centering 
\begin{tabular}{ccccccc} 
	\hline\hline 
	\multicolumn{6}{c}{Particles} \\ [0.9ex]
	\hline 
	&  $N_{\tau R}$ & $N_{\tau L}$ & $N_{ eR}$ & $N_{eL}$ & $\phi$ & $\psi $  \\
	&  $0$ & $1$ & $0$ & $-1$ &$1$& $\frac{g_\psi}{g_N}$ \\
	\hline 
\end{tabular}
\label{tab:charges2}
\end{table}

Let us first discuss how $V_\mu$ can couple to $\nu_\tau$. The coupling to $\nu_e$ or $\nu_\mu$ can be obtained similarly with the relevant replacements. With $N_{\tau L}$ charged under $U_{NEW}(1)$ and mixed with $\nu_\tau$, $V$ can obtain couplings to $\nu_\tau$ as we demonstrate below. Since we reserve $U_{\alpha 4}$ for the mixing between $\nu_\alpha$ and $N_\alpha$ (with $\alpha=e$ or $\mu$), we shall denote the mixing between $N_\tau$ and $\nu_\tau$ by $U_{\tau 5}$.
To obtain a sizable $g_{\nu N}$ coupling, we need both  a large value for $|U_{\tau 5}|$  and a large gauge coupling.
Unless some fine tuned cancellation is at work, taking $N_{\tau}$ as a Majorana fermion and writing a type I seesaw interaction for it, the induced mass for the light active neutrinos would be of order of $m_{N_\tau}   |U_{\tau 5}|^2\gg 1$~eV.  To avoid such large mass induction, we take $N_\tau$ as a Dirac fermion whose mass terms after gauge symmetry breaking are the following
\begin{equation} \label{MAssafterVEV}
\mathcal{L}_m = -\bar{\nu}_\tau m_D N_{\tau R}-M\bar{N}_{\tau L} N_{\tau R}.
\end{equation}

This Lagrangian 
has three mass eigenstates $\tilde{\nu}_\tau$, $\tilde{N}_{\tau R}$ and $\tilde{N}_{\tau L}$ with mass eigenvalues respectively equal to\footnote{ If we added a small Majorana mass for $N_{ \tau L}$ ({\it i.e.,} $\mu_{  L} N_{ \tau L}^T cN_{\tau L}$), we would obtain an inverse seesaw mechanism with an induced mass for $\nu_\tau$ suppressed by the small mass parameter, $\mu_L$.} $0$, $-\sqrt{M^2+m_D^2}$ and $\sqrt{M^2+m_D^2}$ and  the following mixing
\begin{equation} \left(
\begin{matrix} 
	\nu_\tau \cr N_{ \tau R}^c \cr N_{ \tau L}
\end{matrix} \right)=V\cdot \left(
\begin{matrix} 
	\tilde{\nu}_\tau \cr \tilde{N}_{ \tau R}^c \cr \tilde{N}_{\tau L}
\end{matrix} \right)\end{equation}
in which

\begin{equation} V=\left(
\begin{matrix} 
	\frac{M}{\sqrt{M^2+m_D^2}}  & \frac{m_D}{\sqrt{2(M^2+m_D^2)}}  &  \frac{m_D}{\sqrt{2(M^2+m_D^2)}}\cr 0 & \frac{1}{\sqrt{2}}&  \frac{-1}{\sqrt{2}} \cr  \frac{-m_D}{\sqrt{(M^2+m_D^2)}} &  
	\frac{M}{\sqrt{2(M^2+m_D^2)}}	&  \frac{M}{\sqrt{2(M^2+m_D^2)}}				
\end{matrix} \right)\ .\end{equation}
We can therefore write
\begin{equation}
g_NV_\mu \bar{N}_{ \tau L} \gamma^\mu N_{ \tau L}=g_NV_\mu[\sin^2\theta_V \bar{\tilde{\nu}}_\tau \gamma^\mu \tilde{\nu}_\tau+\sin\theta_V\cos\theta_V(\bar{\tilde{\nu}}_\tau \gamma^\mu \tilde{N}_{\tau}+\bar{\tilde{N}}_{\tau} \gamma^\mu \tilde{\nu}_\tau)+\cos^2\theta_V \bar{\tilde{N}}_{\tau}\gamma^\mu \tilde{N}_{\tau} ]
\end{equation}
in which $\tilde{N}_{\tau}=(\tilde{N}_{ \tau L}+\tilde{N}^c_{\tau R})/\sqrt{2}$, $\cos \theta_V=M/\sqrt{M^2+m_D^2}$ and 
$$U_{\tau 5}\equiv \sin \theta_V=\frac{-m_D}{\sqrt{M^2+m_D^2}}\ll 1 ,$$
where $U$ is the full $5 \times 5$ unitary matrix that describes the mixing of the active neutrino with each other and with the new mass eigenstates $N_4$ and $N_5$. The $3\times 3$ PMNS matrix, which now deviates from unitarity,  constitutes a submatrix of $U$.
We can then read the couplings in Eq (\ref{Lag-eff}) as $g_{\nu N}\simeq g_N U_{\tau 5}$ and $g_{\nu \nu}=g_N|U_{\tau 5}|^2$.
There are strong bounds on the $\nu_\tau$ mixing with HNLs from various experiments, such as NOMAD \cite{NOMAD:2001eyx}, CHARM  \cite{Orloff:2002de}, DELPHI \cite{DELPHI:1996qcc} and  ArgoNeut \cite{ArgoNeuT:2021clc},  which come from the search for visible charged particles from the electroweak decay of $N_{\tau}$. In our model, $N_{\tau}$ decays much faster to an invisible mode,  {\it i.e.,} to $\nu_\tau \bar{\nu}_\tau \nu_\tau$. The total number of the $N_\tau$-induced events in the above experiments is given by the $N_\tau$ production rate, which is proportional to $|U_{\tau 5}|^2$, times the branching ratio of the visible decay mode, $N_\tau\to \nu_\tau e^-e^+,\nu_\tau \mu^-\mu^+$. Thus, the bounds from CHARM, NOMAD, DELPHI and ArgoNeut can be relaxed  by the root of the branching ratio of the visible decay, $[{\rm Br}(N_\tau \to{\rm visible)}]^{1/2}\sim \mathcal{O}(G_F m_{N_\tau} m_V/(g_N\pi))\sim 10^{-7}$. However, for $m_{N_{\tau}}<m_\tau$, the $U_{\tau 5}$ coupling is constrained by studying the kinematics of the tau decay modes \cite{BaBar:2022cqj}.
As shown in Fig 10 of \cite{BaBar:2022cqj} by BABAR collaboration,
for $m_{N_{\tau}}<300$ MeV, $|U_{\tau 5}|^2$ can be as large as few times 0.01.
All the relevant bounds for $m_{N_\tau}<$GeV are depicted in Fig 10 of \cite{BaBar:2022cqj}, which, apart from the bounds from the tau decay, come  from CHARM, NOMAD and ArgoNeut under the assumption that $N_\tau$  dominantly decays through its mixing with $\nu_\tau$ via the weak couplings. The strongest bound on $|U_{\tau 5}|^2$, coming from the CHARM experiment \cite{Orloff:2002de},  is $\sim 10^{-4}$. Considering that in the presence of $N_\tau\to \nu_\tau V$ and subsequently $V\to \nu \bar{\nu}$, the bound on $|U_{\tau 5}|^2$ is relaxed by Br($N_\tau\to$visibles), this bound becomes irrelevant. For $m_{N_\tau}<1$ GeV, the so-called unitarity bounds which come from the decay of the states  much heavier than $N_\tau$, such as the $Z$ or $W$ boson, are relaxed. For example, the deficit due to the mixing in the invisible decay mode of $Z\to \nu_\tau \bar{\nu}_\tau$ will be compensated by the new invisible decay modes $Z\to N_\tau \bar{\nu}_\tau,\nu_\tau \bar{N}_\tau,N_\tau \bar{N}_\tau$.

 Thanks to the $g_{\nu \nu}$ coupling, $V_\mu$ can also contribute to the invisible $Z$ boson decay rate as $Z\to V\nu_\tau \bar{\nu}_\tau$. The bound from the $Z$ invisible decay mode  on $g_{\nu \nu}=g_N|U_{\tau 5}|^2$ is around $0.1$ \cite{Bilenky:1992xn,Berryman:2022hds}  which can readily be satisfied for $g_N<1$ and $|U_{\tau 5}|^2<0.01$.  There are strong bounds on the couplings of $\nu_\mu$ and $\nu_e$ from the rare meson decay \cite{Bakhti:2016prn} but since in our model, $V$ couples only to $\nu_\tau$, those bounds do not apply to our model.
As discussed above, the so-called unitarity bound on the mixing of $\nu_\tau$ with HNL cited in \cite{Fernandez-Martinez:2016lgt} assumes an HNL heavier than $\tau$ so it does not apply for $N_\tau$ lighter than 300 MeV.

Similarly, we can couple $\nu_e$ (or $\nu_\mu$) to $V$, by replacing $N_\tau$ with $N_e$ (or  with $N_\mu$), $M_D,M$ with  $M_D^\prime ,M^\prime$ and $U_{\tau 5}, U_{N 5}$ with   $U_{e 4}, U_{N 4}$ (or with $U_{\mu 4}, U_{N 4}$).  In order to have relatively large couplings, we should take $N_\mu$ or $N_e$ heavier than the Kaon to avoid the strong bounds on the relevant mixings from the Kaon decay \cite{NA62:2020mcv}.
However, the unitarity bounds still applies \cite{Fernandez-Martinez:2016lgt}: $|U_{\mu 4}|^2<4.4\times 10^{-4}$ and $|U_{e 4}|^2<5.6\times 10^{-3}$.

Let us now discuss how Eq.~(\ref{MAssafterVEV}) can be obtained from a gauge invariant Lagrangian. Without attributing $U_{NEW}(1)$ charges to the SM particles, $m_D$ breaks the gauge symmetry unless $N_{\tau R}$ is a gauge singlet. Nonzero $M$ then requires gauge symmetry breaking which can be achieved by  the  vacuum expectation value of a new scalar $\phi$ with a $U_{NEW}(1)$ charge equal to that of $N_{
\tau L}$. With  this field content, Eq. (\ref{MAssafterVEV}) can be obtained after  gauge symmetry breaking from
\begin{equation} \label{NLcoup} 
Y_{\nu N} \bar{N}_{\tau R} H^TcL_\tau-\lambda_{LR} \bar{N}_{\tau L} N_{\tau R} \phi .
\end{equation}
Then, $m_D = Y_{\nu N}\langle H \rangle$, $M =\lambda_{LR} \langle \phi\rangle$.
Since the $N_\tau$ interaction is chiral, it can induce a $(U_{NEW}(1))^3$ anomaly
which will be canceled by the contribution from $N_{eL}$ (or from $N_{\mu L}$) with an opposite $U_{NEW}(1)$ charge. We can also write Yukawa couplings for $N_{eL}$ (or $N_{\mu L}$), replacing $Y_{\nu N} \to  Y_{\nu N}^\prime$, $\lambda_{LR} \to \lambda_{LR}^\prime$ and $\phi$ with $\phi^*$.  
We stick to the lepton flavor conserving case where each HNL mixes with (at most) one active neutrino flavor.

Let us now discuss the Lagrangian of  $\psi$. Since we have taken $\psi$ to be Dirac, the mass term can be explicitly written as $m_\psi \bar{\psi}\psi$. If we take $\psi$ to be chiral, such that  $e.g.$ only $\psi_L$ is charged under $U_{NEW}(1)$ with a charge opposite to that of $N_{ \tau L}$,  the $U_{NEW}(1)^3$ anomaly can be canceled with only one $N_{ \tau L}$ without any need for $N_{e L}$ (or $N_{\mu L}$). Then, we could use $\phi$ to write a Yukawa coupling that after  $U_{NEW}(1)$ gauge symmetry breaking would lead to a mass for $\psi$. Within this minimalistic model, the neutrino and antineutrino cross sections would differ. To avoid this complication, we have taken $\psi$ to be non-chiral.  In this case, the  $U_{NEW}(1)$ charges of $N_{\tau L}$ and $\psi$ can be independent. 

Being a Dirac fermion, $\psi$  can fit within asymmetric dark matter scenarios  \cite{Nussinov:1985xr,Barr:1990ca,Barr:1991qn,Kaplan:1991ah,Davoudiasl:2012uw,Zurek:2013wia, Dev:2026ddq}. Both $\psi$ and $\bar{\psi} $ can be thermalized in the early universe when the temperature of the plasma is higher than the $\psi$ mass, $m_\psi$. When the temperature of the plasma drops below $m_\psi$, the $\psi \bar{\psi}$ annihilation with a cross section of $$ \sigma(\psi\bar{\psi}\to \nu \bar{\nu})\sim \frac{g_\psi^2 g_N^2|U_{\tau 5} |^4}{4\pi m_\psi^2}=3 \times 10^{-31}~{\rm cm}^2 g_\psi^2 g_N^2 \frac{(0.1~{\rm GeV})^2}{m_\psi^2} \frac{|U_{\tau 5}|^4}{10^{-4}}\ll  1~{pb}\ ,$$ will reduce the $\bar\psi$ abundance to  a freeze-out value $\Omega_{\bar{\psi} }/\Omega_{DM} \sim (1~pb)/\sigma\sim 10^{-5}$. Thus, within the asymmetric dark matter scenario composed of ${\psi}$, we can safely neglect the presence of $\bar{\psi}$, today. If the spike is made of only $\psi$ (without a $\bar{\psi}$ contribution), dark matter annihilation cannot take place so the spike profile can be extended down to very large values at $R_{em}$.
We have also developed a coannihilation  dark matter production scenario, with a pair of chiral Majorana fermions, $\chi_L$ and $\chi_L^\prime$ with  $m_{\chi^\prime}> m_\chi$ and a coupling of form $( \bar{\chi}_L^\prime \gamma^\mu \chi_L)V_\mu$. 
$\chi^\prime$ can decay to $\chi \bar{\nu}_\tau \nu_\tau$ so its relic density would be too small to lead to coannihilation in the spike.
We however noticed that  even if all new particles of the model are heavier than $2m_\chi$, $\chi \bar{\chi}$ can still annihilate to two $\nu_\tau \bar{\nu}_\tau$ pairs with  the following cross section 
$$ \sigma(\chi {\chi} \to \nu_\tau \bar{\nu}_\tau  \nu_\tau \bar{\nu}_\tau )\sim \frac{|g_N|^4|g_\chi|^4 }{4\pi(2 m_\chi)^2} \frac{|U_{\tau 5}|^8}{(16 \pi^2)^2} \stackrel{<}{\sim} 10^{-40}~{\rm cm}^2.$$  Even if we massage the model to prevent the $\bar{\nu} \gamma^\mu \nu V_\mu$ coupling and therefore $\chi \chi \to \nu \nu\bar{\nu}\bar{\nu}$, at one loop level, we shall still have a $p$-wave annihilation to $\nu_\tau \bar{\nu}_\tau$:
$$ \sigma(\chi {\chi} \to \nu_\tau \bar{\nu}_\tau  )\sim \frac{|g_N|^4|g_\chi|^4 }{4\pi(2 m_\chi)^2} \frac{|U_{\tau 5}|^4}{(16 \pi^2)^2}p_\chi,$$ 
where $p_\chi$ is the momentum of the annihilating $\chi$ at the spike. At $R_{em}$, $p_\chi$ can be estimated as $p_\chi\sim (R_S/2R_{em})^{1/2}\sim 1/\sqrt{60}$.
Although these annihilation cross sections seem to be small but corresponds to saturation densities \cite{Vasiliev:2007vh,Shapiro:2016ypb} far smaller than that of NGC 1068 at $R_{em}$  and therefore to a suppressed and negligible $\Sigma$.  As a result, we shall not further elaborate on this scenario and will instead assume  that dark matter is asymmetric.

We are interested in the part of the parameter space where the cross section of the $\nu_\tau$ scattering  off $\psi$ is sizable such that
\begin{equation} \tau_{SP}\equiv \frac{\Sigma\sigma} {m_\psi} > 1.
\end{equation}
This in turn constrains the gauge boson mass.  Using Eq.~(\ref{eq:scatterX}), we can write
\begin{equation}
m_V=390~{\rm MeV} g_\psi g_N\left(  \frac{ 10^{-30}~{\rm cm}^2}{\sigma} \frac{|U_{\tau 5}|^2}{0.01} \right)^{1/2} \  . \label{mVfromSigma}
\end{equation}
Considering that $m_V$,  $m_{N_\tau}$ and $m_{N_{e(\mu)}}$ are all determined by $\langle \phi \rangle$,  Eq.~(\ref{mVfromSigma}) implies
\begin{equation}
m_{N_\tau}=270~{\rm MeV} g_\psi \frac{\lambda_{LR} }{0.7}\left( \frac{ 10^{-30}~{\rm cm}^2}{\sigma}\frac{|U_{\tau 5}|^2}{0.01} \right)^{1/2} \  \label{mNfromSigma}
\end{equation}
and
\begin{equation}
m_{N_{e(\mu)}}=500~{\rm MeV} g_\psi \frac{\lambda_{LR}^{\prime} }{1.3}\left( \frac{ 10^{-30}~{\rm cm}^2}{\sigma}\frac{|U_{\tau 5}|^2}{0.01} \right)^{1/2} \  . \label{mNfromSigma}
\end{equation}

\begin{figure*}
\centering
\subfloat[]{\includegraphics[width=0.49\textwidth ]{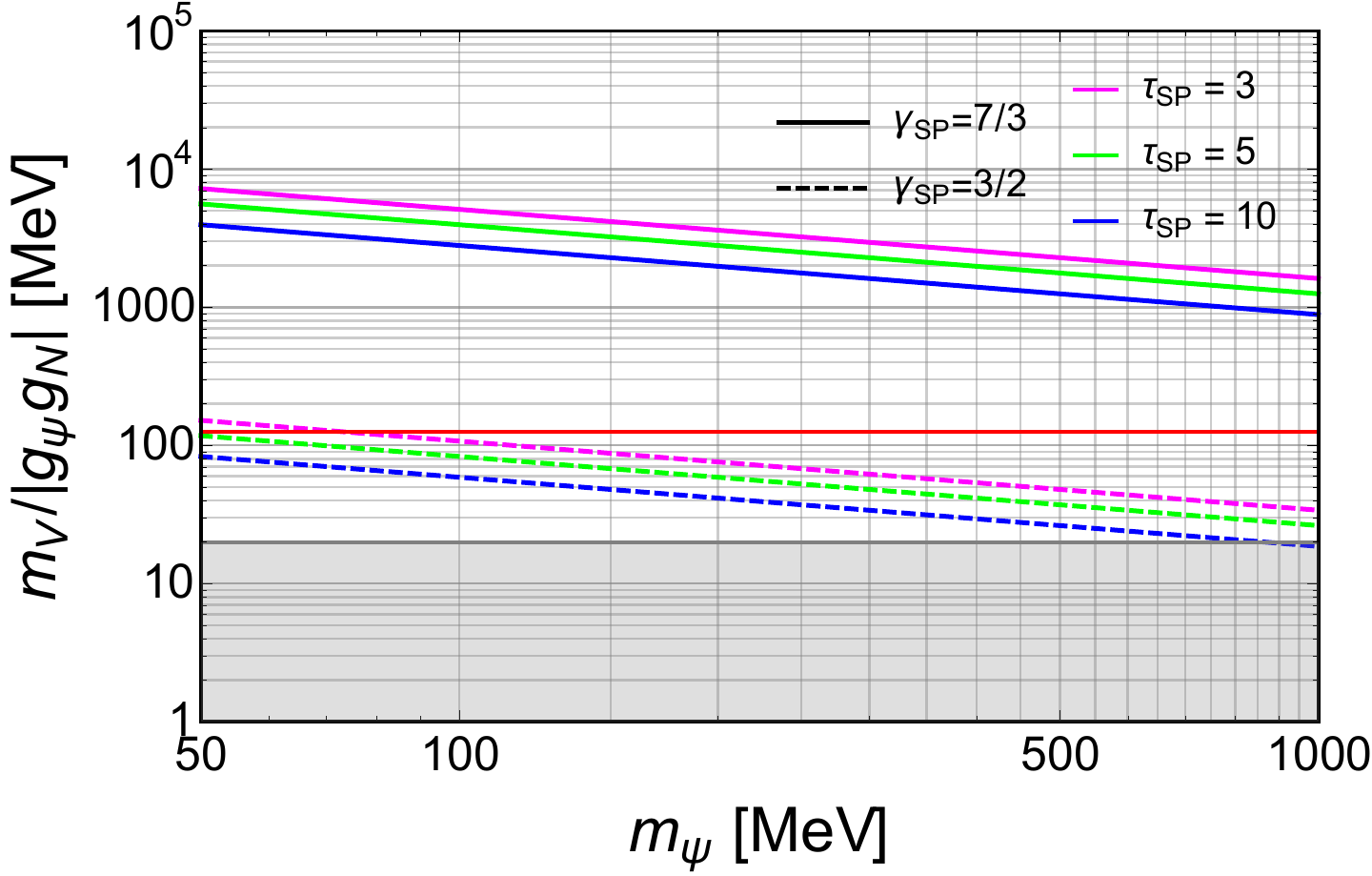}\label{fig:Mv-mchi-tausp}}
\subfloat[]{\includegraphics[width=0.49\textwidth ]{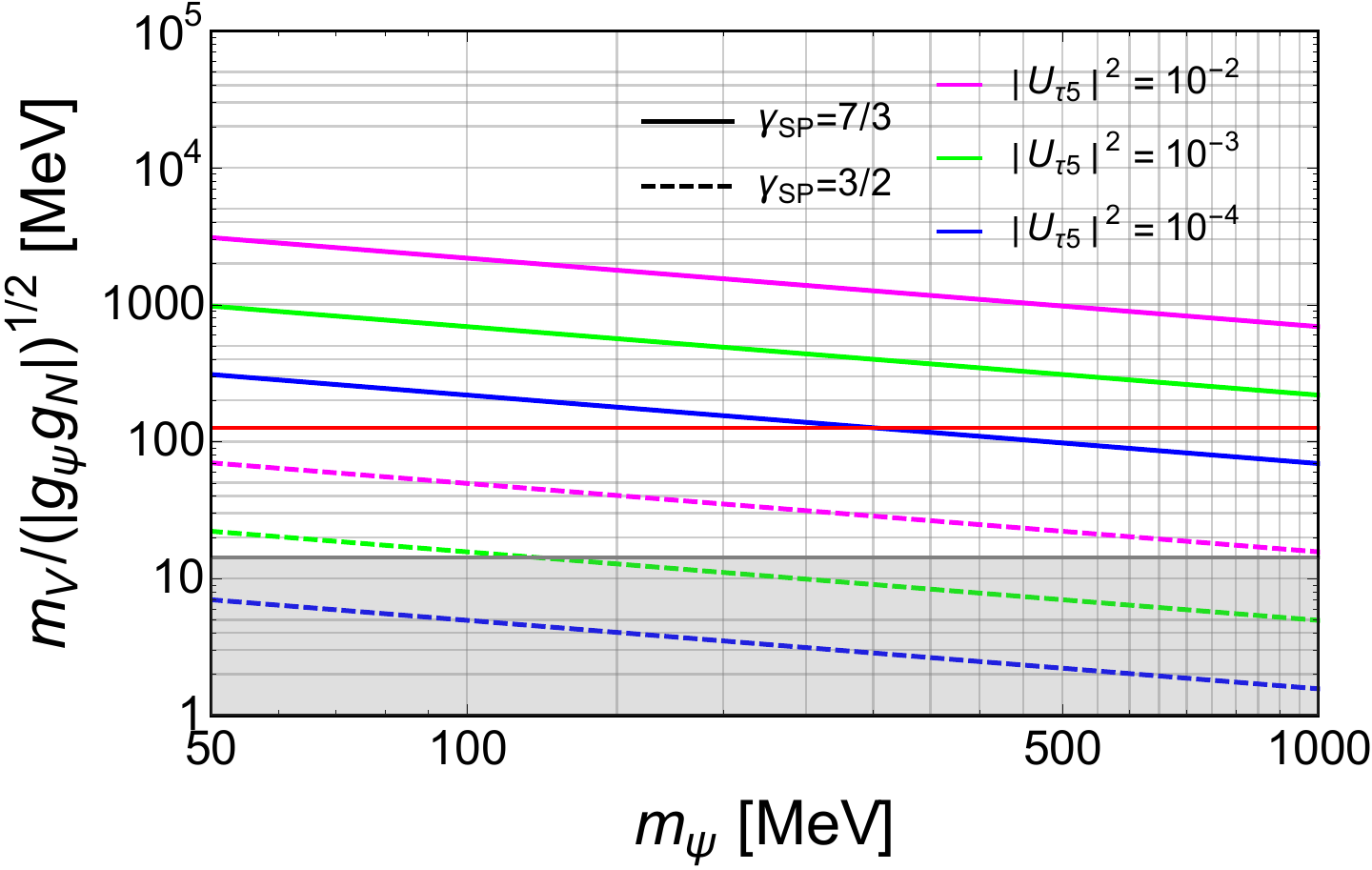}\label{fig:Mv-mchi-Atautau}}
\caption{Left : The magenta, green and dark blue lines  show  the values of $m_V/|g_N g_\psi|$ versus $m_\psi$ that respectively  lead to  $\tau_{SP}=3,5$ and $10$ in the NGC 1068 spike for $|U_{\tau 5}|^2=10^{-2}$. Right: The same color scheme is used to show the values of $m_V/\sqrt{|g_N g_\psi|}$ versus $m_\psi$  corresponding to $\mathcal{V}_\tau=10^{-16}$~eV  (see Eqs. (\ref{Hcal},\ref{Ecal})) for $|U_{\tau 5}|^2=10^{-2}, 10^{-3}, 10^{-4}$, respectively. In both panels, the solid and dashed lines correspond to  $\gamma_{SP}=7/3$ with $\rho_{spike}(R_{em})=6.24\times10^{17}~ \mathrm{GeV/cm^3}$  and $\gamma_{SP}=3/2$ with $\rho_{spike}(R_{em})=3.2\times10^{14} ~\mathrm{GeV/cm^3}$, respectively. The gray area is disfavored by a combination of the lower bound on $m_V$ from the $N_{eff}$ measurement and the coupling perturbativity. Below the red  horizontal line $m_{N_{e(\mu)}}$ cannot be larger than the Kaon mass  with $g_\psi$ and $\lambda_{LR}^\prime$ in the  perturbative  range. }
\label{mVsigma}
\end{figure*}

In our model, reducing the value of $\lambda_{LR}$, light $N_\tau$ can be achieved but on the other hand, $N_\tau$ should be heavier than $\sim 20$ MeV so that it can decay away before the neutrino decoupling era. Otherwise, $\nu_\tau \bar{\nu}_\tau \nu_\tau$ can increase the effective relativistic degrees of freedom, $N_{eff}$ which is constrained by the BBN and CMB data.  For similar reasons, $V$ should also be heavier than $\sim 20$ MeV.    Taking $| g_\psi g_N|<1$, this means $m_V/|g_\psi g_N|>20$~MeV which is shown with the gray region in Fig.~\ref{mVsigma}. 
Taking $m_V<m_{N_\tau}$, the  decay rate of $N_\tau$ is given by \cite{Liu:2025akp}
$$
\Gamma(N_\tau \to\nu_\tau V)=
\frac{g_N^2|U_{\tau 5}|^2}{32\pi}
\frac{m_{N_\tau }^3}{m_V^2}
\left(1-\frac{m_V^2}{m_{N_\tau }^2}\right)^2
\left(1+2\frac{m_V^2}{m_{N_\tau }^2}\right).
$$ Then, $V$ can decay to a pair of neutrinos with
$$
\Gamma(V\to\nu_\tau \bar\nu_\tau )
=\frac{g_{N}^2|U_{\tau 5}|^4 m_V}{24\pi} .$$
It is straightforward to show that for $20~{\rm MeV}<m_V<m_{N_{\tau}}$ and for the range of parameters of interest in this paper ($g_N \sim 1$ and $|U_{\tau 5}|^2>10^{-5}$), the decay rates of $V$ and  $N_{\tau}$ are much faster than the Hubble expansion rate at $T=m_V$ or at  $T=m_{N_\tau}$,  implying that at the time of decoupling, their contributions to the energy budget of the Universe are suppressed by the Boltzmann factor $\exp (-m_V/T)\ll 1$ or 
$\exp (-m_{N_\tau}/T)\ll 1$, respectively. For a more comprehensive study of the cosmological bounds, see Ref \cite{Dev:2025tdv}.
In the early universe, $\psi \bar{\psi}$ reach thermal equilibrium. Then, $\psi \bar{\psi} \to \nu_\tau \bar{\nu}_\tau$ along with other possible annihilation modes  severely deplete the relic $\bar{\psi}$ abundance before temperature $\sim m_\psi/20$.  However, the asymmetric $\psi$ component can survive.  Again, the consideration from the upper bound on $N_{eff}$ implies $m_\psi \stackrel{>}{\sim} 20$~MeV. To be on the safe side, we take $m_\psi$ heavier than 50~MeV.
On the other hand, to avoid the bound found in Refs.~\cite{Chauhan:2025hoz,Esteban:2025wbv,Bertolez-Martinez:2025trs}, we take $m_\psi < 1$~GeV. 
Fig.~\ref{fig:Mv-mchi-tausp} shows the value of $m_V/|g_N g_\psi|$ (corresponding to a given value of $\sigma$; see Eq. (\ref{mVfromSigma})) versus $m_\psi$ leading to $\tau_{SP}=3,5,10$ in the spike of an AGN such as NGC 1068 with  $\gamma_{SP}=7/3$ (solid lines) and with  $\gamma_{SP}=3/2$ (dashed lines). The gray region is ruled out by a combination of perturbativity of the couplings and negligible contribution to $N_{eff}$ as explained above. As seen from the figure, with $\gamma_{SP}=7/3$, $\tau_{SP}=10$ can be achieved for the entire range $50~{\rm MeV}<m_\psi<1~{\rm GeV}$ . With $E_\nu<$PeV and $\gamma_{SP}=3/2$,  $\tau_{SP}=10$ can also be obtained as long as $m_\psi<800$~MeV. 
Below the red line, $N_{e(\mu)}$ cannot be heavier than the  Kaon so its mixing should be small. Thus, from Fig~\ref{mVsigma}, we conclude that if $\gamma_{SP}=3/2$, new physics can appear  only  as a consequence of  the $\nu_\tau$ mixing. However, as we shall see in sect. \ref{medium}, through the forward scattering effects, this can  still severely affect the flavor ratios of the high energy neutrino fluxes reaching the Earth through the spikes of AGNs.

\section{ Flavor measurement by IceCube and IceCube-Gen2 \label{ice}}

During over a decade of data taking, IceCube not only has observed more than hundred  high energy astrophysical neutrinos but has also determined their flavor composition with a precision that was not even imaginable in the beginning of the IceCube PMT deployment.  The dark blue contours  shown in  Figs.~(\ref{fig:ternery-Vt-Standard-ratio},\ref{fig:ternery-Ve-Vt-lowenerg},\ref{fig:ternery-Vmu-Vt-lowenerg}) represent the most recent IceCube results based on studying 11.4 years of IceCube data on the cosmic neutrino flux in the energy range from 5 TeV to 10 PeV \cite{Abbasi:2025fjc}. These contours are derived assuming a broken power-law (BPL) spectrum for the astrophysical neutrino flux in this energy range. IceCube has also performed a fit assuming a single power law (SPL) spectrum, obtaining a best fit flavor composition of $(\nu_e^\oplus:\nu_\mu^\oplus:\nu_\tau^\oplus)$= $(0.28, 0.36, 0.36)$, which is very close to the BPL best fit value of $(0.30, 0.37, 0.33)$.
As seen from the figure, while the standard prediction for the flavor ratios at the source $(\nu_e^S:\nu_\mu^S:\nu_\tau^S)=(1/3,2/3,0)$ (or equivalently that at the Earth 
$(\nu_e^\oplus:\nu_\mu^\oplus:\nu_\tau^\oplus)\simeq (1/3,1/3,1/3)$) is compatible with the measurement,  at 1~$\sigma$ C.L., the flavor ratio measurement already disfavors the neutron decay as the source of the high energy neutrino flux  for which 
$(\nu_e^S:\nu_\mu^S:\nu_\tau^S)=(1,0,0)$.

IceCube-Gen2 is going to be an extension of IceCube with an instrumented volume eight times larger than IceCube  whose construction will be complete around 2033 \cite{IceCube-Gen2:2020qha}.  It will be able to determine the flavor ratios with better precision. For forecasting the uncertainty of the flavor ratio determination by 
IceCube-Gen2, three energy ranges are distinguished: (1) $E_\nu<$PeV; The $1 \sigma$ and $2\sigma$ forecast of Ref. \cite{IceCube-Gen2:2020qha} under the canonical  assumption of $(1/3,2/3,0)$ is shown in Fig.~\ref{fig:ternery-Vt-Standard-ratio}-\ref{fig:ternery-Vmu-Vt-lowenerg}  with black contours; (2) ${\rm PeV}<E_\nu<100$ PeV for which  a muon damped source with the initial flavor ratio  $(0,1,0)$ is assumed \cite{IceCube-Gen2:2025yte}.  In Fig.~\ref{fig:ternaryHIGH}, the contours show the forecast for IceCube-Gen2. (3)  $E_\nu >100$ PeV for which the neutrino source  is expected to be  cosmogenic ({\it i.e.}, the scattering of ultra-high energy cosmic ray off CMB) rather than AGN activities.

By the time that IceCube-Gen2 is ready for the  data collection, new ideas and techniques may help to determine the flavor ratios even more precisely than what was anticipated in the original plan. For example, Ref. \cite{Wen:2026ngx} suggests to use ``visible inelasticity" to probe the $\nu_\tau$ content of the flux. Moreover, techniques are being developed that make the separate measurement of the $\nu_\mu$ and $\bar{\nu}_\mu$ fluxes possible \cite{Skrzypek:2025tmg}.

\section{Dark matter  induced matter potential in the medium of the spike \label{medium}}
In a medium, the quantum evolution of the neutrino states (without hard scattering) is governed by the following effective Hamiltonian 
\begin{equation}
\mathcal{H}=U_{PMNS}.{\rm Diag}[\frac{m_1^2}{2E_\nu},\frac{m_2^2}{2E_\nu},\frac{m_3^2}{2E_\nu}]\cdot U_{PMNS}^\dagger+{\rm Diag}[\mathcal{V}_e,\mathcal{V}_\mu,\mathcal{V}_\tau] \ . \label{Hcal}
\end{equation}
The first term is the Hamiltonian in vacuum, $\mathcal{H}^{vac}$, and the second term is induced by the coherent forward scattering off the particles composing the medium. Within the standard model,  this term appears in the medium of dense environments such as stars or planets. Within our model, in the medium of the spike composed of merely $\psi$ (with negligible contribution from $\bar{\psi}$), we can write
\begin{equation}
\mathcal{V}_e=-g_\psi g_N\frac{\rho_{spike}}{m_\psi}\frac{|U_{e4}|^2}{m_V^2} \ , \ \  	\mathcal{V}_\mu=0 \ \ {\rm and}\ \  \mathcal{V}_\tau=g_\psi g_N\frac{\rho_{spike}}{m_\psi}\frac{|U_{\tau 5}|^2}{m_V^2} \label{Ecal}
\end{equation} 
for the case that $U_{\mu 4}=U_{\mu 5}=0$, and
\begin{equation}
\mathcal{V}_e=0  \ , \ \  	\mathcal{V}_\mu=-g_\psi g_N\frac{\rho_{spike}}{m_\psi}\frac{|U_{\mu 4}|^2}{m_V^2} \ \ {\rm and} \ \  \mathcal{V}_\tau=g_\psi g_N\frac{\rho_{spike}}{m_\psi}\frac{|U_{\tau 5}|^2}{m_V^2}
\end{equation} 
for the case that $U_{e 4}=U_{e5}=0$.
The signs of $\mathcal{V}_\tau$ and $\mathcal{V}_e$ ($\mathcal{V}_\mu$) are opposite. This is because to cancel the $(U_{NEW}(1) )^3$ anomaly, opposite charges are assigned to  $N_\tau$  and $N_e$ ($N_\mu$).
The quantum evolutions of the anti-neutrino states are also given by Eq. (\ref{Hcal}) but with an opposite sign for $\mathcal{V}_\alpha$.

The effective Hamiltonian $\mathcal{H}$ at any point across the spike can be diagonalized as 
\begin{equation} \mathcal{H}=U_{SP}\cdot M_{diag}^{SP}\cdot U_{SP}^\dagger
\end{equation}
in which 
$M_{diag}^{SP} ={\rm Diagonal}[\tilde{m}^2_1/(2E_\nu),\tilde{m}^2_2/(2E_\nu),\tilde{m}^2_3/(2E_\nu)]$.
If the variation of $\mathcal{V}_\alpha$ and therefore  those of ${U_{SP}}$ and $
M_{diag}^{SP}$ are slow, the evolution will be adiabatic and the probability of jump between different effective mass eigenstates, $P_J(\tilde{\nu}_i \to \tilde{\nu}_j)$ will be negligible. The flavor transition probabilities for neutrinos propagating from $R_{em}$ to $r$ can then be written as 
\begin{equation}
P(\nu_\alpha \to \nu_\beta)=\left|\sum_i (U_{SP})_{\alpha i} |_{R_{em}}(U_{SP}^*)_{\beta i} |_{r} \exp(i\int_{R_{em}}^r\frac{\tilde{m}^2_i}{2E_\nu}dr')\right|^2 \ .
\end{equation}
At $r\gg R_S$, matter effects are negligible so $U_{SP}=U_{PMNS}$. Moreover, $\int_{R_{em}}^r\frac{\tilde{m}^2_i}{2E_\nu}dr'\gg 1$ so when neutrino exits the spike (as well as when it arrives at the Earth), 
$$ P(\nu_\alpha \to \nu_\beta)=\sum_i |(U_{SP})_{\alpha i}(U_{PMNS}^*)_{\beta i}|^2\ .$$

Let us now examine the adiabaticity of the evolution. For the neutrinos propagating in the spike away from the center,
$$ \left|\frac{d\mathcal{V}_\alpha}{dt}\right|=
c\left|\frac{d\mathcal{V}_\alpha}{dr}\right|
=\gamma_{SP} \left|\frac{\mathcal{V}_\alpha}{r}\right|
\ . $$
Using the standard parametrization for $U_{SP}$ with $\tilde{\theta}_{12}$, $\tilde{\theta}_{23}$ and $\tilde{\theta}_{13}$, the adiabacity condition can be formulated as 
\cite{Kim-Giunti}
$$1 \ll \gamma_{ij}=\frac{\Delta \tilde{m}_{ij}^2}{4 E_\nu d \tilde{\theta}_{ij}/dt} \ . $$
Notice that because of the specific flavor structure of $\mathcal{V}_\alpha$ ({\it i.e.,} $|\mathcal{V}_\tau|>|\mathcal{V}_e|,|\mathcal{V}_\mu|$) and the proximity of $\theta_{23}$ to $\pi/4$, the resonances are not very severe. That is even at the resonance, $\Delta \tilde{m}_{ij}^2/\Delta {m}_{ij}^2 \sim 1$. Moreover, $|d \Delta \mathcal{V}/dr|$ decreases with $r$. As a result, the minimum of $\gamma_{ij}$ is not necessarily located at the resonance and can be around $R_{em}$.
Taking $\mathcal{V}_\alpha(r)=
\mathcal{V}_\alpha(R_{em})(r/R_{em})^{-\gamma_{SP}}$ with $|V_\tau(R_{em})|\gtrsim 10^{-16}$ eV,  we computed $\gamma_{ij}$ within  the parameter range of  our interest.   We find that for $70~{\rm TeV}<E_\nu<100$~PeV, the adiabacity condition is fulfilled across the spike with $\gamma_{ij}>10$ and therefore with a  negligible probability of non-adiabatic transition given by $\exp(-4\pi \gamma_{ij})$.

Ref.~\cite{Farzan:2018pnk} also studies the impact of the dark matter 
induced  matter potential on the flavor ratios of high energy neutrinos reaching the Earth, focusing on   the effect of the halos of the host galaxy as well as that of the Milky Way. In the model presented in  Ref.~\cite{Farzan:2018pnk}, the flavor eigenstates coming out of the source will convert to the mass eigenstates at vacuum ({\it i.e.,} $\nu_e \to \nu_1$, $\nu_\mu \to \nu_2$) and then back to the flavor eigenstates in the Milky Way halo so that the flavor ratio $(\nu_e:\nu_\mu:\nu_\tau)=(1/3:2/3:0)$ would be preserved along the way. In the present study, the dark matter effects are non-negligible only at very high densities inside the spike so that the Milky Way halo will not be relevant. The fluxes of the mass eigenstates exiting the spike will arrive at the Earth as the same mass eigenstates rather than the flavor eigenstates.

Notice that 
\begin{equation} \frac{\rho_{spike}}{m_\psi}\sigma(\nu_\alpha +\psi \to N_\alpha +\psi)=\frac{\left| g_\psi g_N \mathcal{V}_\alpha \right|}{4\pi}.\label{prop} 
\end{equation}
Taking a power law behavior for the spike density profile, $\rho_{spike}\propto r^{-\gamma_{SP}}$, Eq. (\ref{prop}) implies that  the ratio of the optical depth of $\nu_\alpha$ to $\mathcal{V}_{\alpha}$ will be independent of $\rho_{spike}$ at the production. The ratio will be  even independent of the value of $\gamma_{SP}$. In table \ref{tab:benchmark1}, we show the values of the model parameters, along with the optical depths and $\mathcal{V}_\alpha$ at a few representative benchmarks that we shall study. We have fixed $m_\psi=100$ MeV for all these benchmarks. The energy threshold of IceCube-Gen2 will be 70~TeV for which $\Delta m_{31}^2/(2E_\nu)=2\times 10^{-17}$~eV. As a result, for $|\mathcal{V}_\alpha|\gtrsim 10^{-16}$~eV, we are in the regime of  large matter effects.

\begin{table*}[htbp]
\centering
\caption{Benchmark parameters and the corresponding effective potential in matter,  $\mathcal{V}_\alpha$, at $\rho_{spike}(R_{em})=6.24\times 10^{17}$~GeV$/{\rm cm}^3$.  The last two columns show the total flux attenuation with two neutrino energy spectral indexes, $\gamma_\nu=3.2$ and $\gamma_\nu=2.5$.
	The dark matter mass is fixed to $m_{\psi}=100 ~\mathrm{MeV}$ for all the benchmark points and the spike profile is determined with $\gamma_{SP}=7/3$. Varying $|U_{\tau 5}|^2$ in the indicated ranges, $\mathcal{V }_\tau$ and the attenuation will vary in the ranges shown in the table; however, the predicted final flavor ratios will remain almost constant.}
\label{tab:benchmark1}
\setlength{\tabcolsep}{0.4pt}
\renewcommand{\arraystretch}{1.1}
\rowcolors{1}{gray!10}{cyan!10}
\begin{tabular}{cccccccccccc}
	\hline\hline
	& $g_{\psi}$ & $g_N$ &  $\frac{m_{V}}{\rm{MeV}}$ &$|U_{e4}|^2$& $|U_{\mu 4}|^2$& $|U_{\tau5}|^2$ &$\frac{\mathcal{V}_{e}}{10^{-16}\rm eV}$ & $\frac{\mathcal{V}_{\mu}}{10^{-16}\rm eV}$& $\frac{\mathcal{V}_{\tau}}{10^{-16}\rm eV}$&\makecell{Attenuation\\($\gamma_\nu=3.2$)} &  \makecell{Attenuation\\($\gamma_\nu=2.5$)}  \\
	\hline\hline
	
	$\rm{B}_1$ &0.2 &0.27 &120 &$\frac{10^{-3}}{\sqrt{3}}$ &- & $(10^{-3},10^{-2})$  &-1.03& -& $(1.8,18)$ &(0.74,0.78)&(0.77,0.86) \\
	\hline
	$\rm{B}_2$ &-0.2 & 0.27& 120&$\frac{10^{-3}}{\sqrt{3}}$ &-&$(10^{-3},10^{-2})$ &  1.03&- &(-18,-1.8)&(0.74,0.78)&(0.77,0.86)\\
	\hline
	$\rm{B}_3$ &0.4 &0.2 &60 &-&$10^{-4}$ &$(10^{-4},10^{-2})$ &- &-1.06 & $(1.06,106)$&(0.47,0.58)&(0.52,0.79) \\
	\hline
	$\rm{B}_4$ &-0.4 &0.2 &60 &-&$10^{-4}$&$(10^{-4},10^{-2}) $  &- &1.06 &$(-106,-1.06)$&(0.47,0.58)&(0.52,0.79)\\
	\hline
	\hline
\end{tabular}
\end{table*}

For $|\mathcal{V}_\alpha|\ll 10^{-16}$ eV, not only the matter effects at $E_\nu\sim100$~TeV are suppressed but the scattering rate off the spike will also be small  (see Fig. \ref{mVsigma}) so the presence of the spike will be irrelevant. On the other hand, it seems quite an unlikely coincidence that $|\mathcal{V}_\alpha(R_{em})| \sim \Delta m_{ij}^2/E_\nu$ at $E_\nu \sim 100$ TeV. 
Moreover, in this case the final flavor ratio will be very sensitive to $\rho_{spike}(R_{em})$ which is unknown and may differ from AGN to AGN, undermining the predictivity. 
Dismissing these two cases, we shall focus on $|\mathcal{V}_\tau|> 10^{-16}$ eV at $R_{em}$ and  the parts of the parameter space  with  robust predictions for the flavor ratios.

The relation between the scattering rate $\rho_{spike}\sigma /m_\psi$ and $\mathcal{V}_\alpha$ shown in Eq.~(\ref{prop}) helps us to considerably simplify an otherwise very complicated computation of the flux evolution in the presence of both  the inelastic scattering and the flavor conversion due to the mixing. The scattering is relevant only in the dense regions where $\mathcal{V}_\alpha$ is high, leading to a suppressed effective flavor mixing. Thus, we can use the cascade equations shown in appendix \ref{cascade} without worrying about the flavor evolution due to the mixing. 

\begin{figure*}[!ht]
\centering
\includegraphics[width=1\textwidth ]{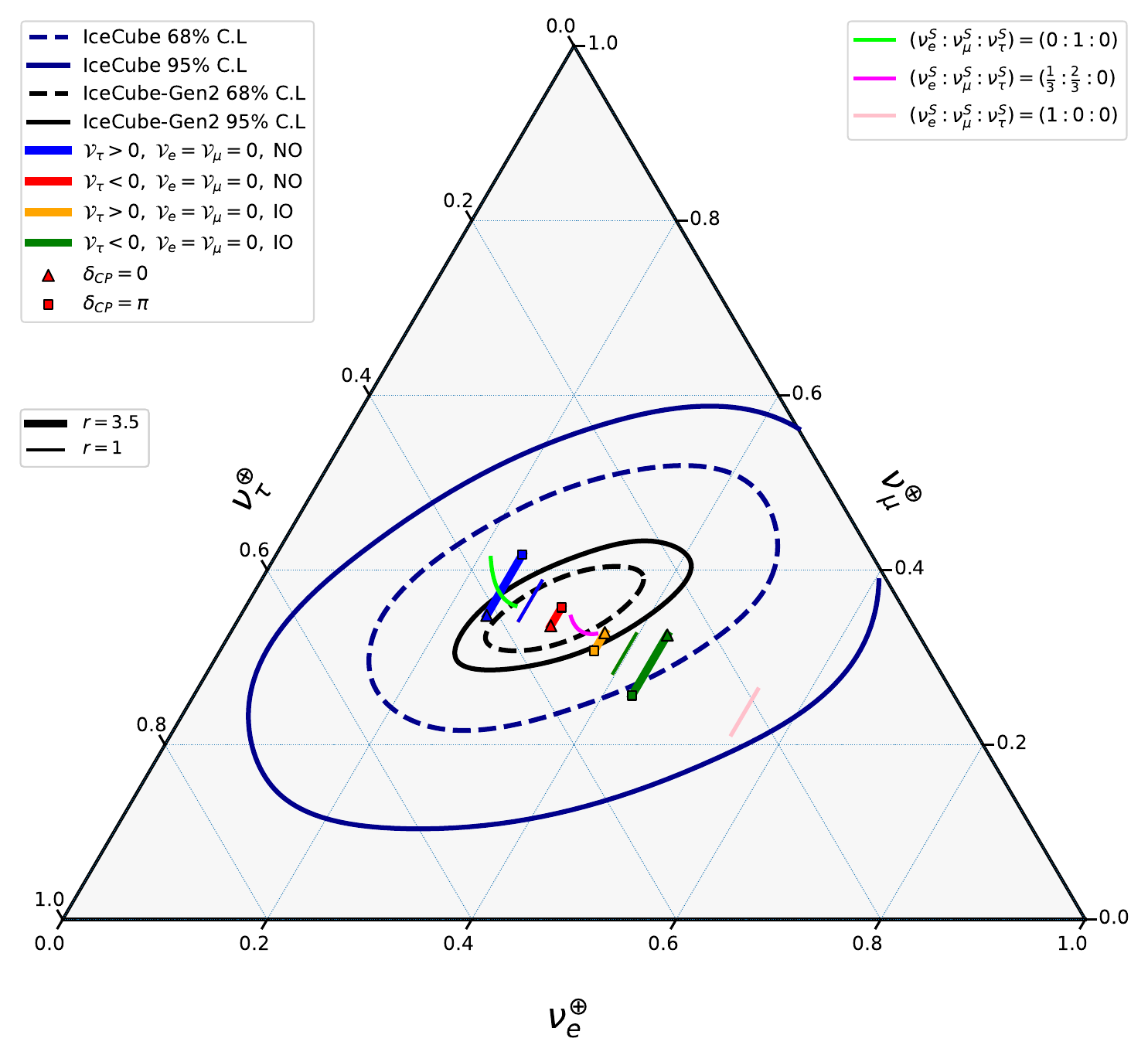}
\caption{Flavor ratios at the Earth for  $\mathcal{V}_e=\mathcal{V}_\mu=0$ and $
	|\mathcal{V}_\tau|> 10^{-16}$ eV,  starting with  $\Phi^S_{\bar{\nu}_e}=r\Phi^S_{\nu_e}$ and  $\Phi^S_{\nu_\mu}=\Phi^S_{\bar{\nu}_\mu}=\Phi^S_{\nu_e}+\Phi^S_{\bar{\nu}_e}$ at the production and varying $0<\delta_{CP}<2\pi$. Thick and thin line segments correspond to $r\equiv\Phi^S_{\bar\nu_e}/\Phi^S_{\nu_e}=3.5$ and $1$, respectively. The dark blue, red, orange, and  green segments respectively show our predictions for $\mathcal{V}_\tau>0$ with $\Delta m_{31}^2>0$,
	$\mathcal{V}_\tau<0$ with $\Delta m_{31}^2>0$, $\mathcal{V}_\tau>0$ with  $\Delta m_{31}^2<0$,  and $\mathcal{V}_\tau<0 $ with $\Delta m_{31}^2<0$. 
	The  latest IceCube measurement \cite{Abbasi:2025fjc} and IceCube-Gen2 forecast  \cite{IceCube-Gen2:2025yte} are respectively shown by the dark blue and black contours.  
	For comparison, we have shown the standard prediction without scattering for  different flavor ratios at the source with colored segments (see the top right legend box).    
}
\label{fig:ternery-Vt-Standard-ratio}
\end{figure*}

As we saw in sect. \ref{model}, from the theoretical point of view, a flavor structure  of $\mathcal{V}_e=\mathcal{V}_\mu=0$ and $
\mathcal{V}_\tau\ne 0$ is motivated.
If only $|\mathcal{V}_\tau|$ is large, we can write  the flavor eigenstates in terms of the effective mass eigenstates as
\begin{eqnarray}
|\nu_e\rangle &=& \cos \theta|\tilde{\nu}_1\rangle+ \sin \theta|\tilde{\nu}_2\rangle \cr
|\nu_\mu\rangle &=& -\sin \theta|\tilde{\nu}_1\rangle+ \cos \theta|\tilde{\nu}_2\rangle\cr
|\nu_\tau\rangle &=& |\tilde{\nu}_3\rangle
\end{eqnarray}
where $\tan 2 \theta\simeq \mathcal{H}^{vac}_{12}/\mathcal{H}^{vac}_{22}\simeq 0.19$. The formulas are different from the standard familiar case because while the standard matter effect leads to $\mathcal{V}_e\ne \mathcal{V}_\mu=\mathcal{V}_\tau$, in this case, $\mathcal{V}_e= \mathcal{V}_\mu\ne \mathcal{V}_\tau$. For zero mixing  of  HNL with $\nu_e$ and $\nu_\mu$, the fluxes of these neutrinos will not suffer from attenuation. The flavor ratios of neutrinos at Earth will however depend on both mass ordering of ordinary neutrinos and the sign of $\mathcal{V}_\tau$. Thus, the following four cases can be distinguished  for neutrinos. (We shall comment on the flavor composition of the antineutrinos with opposite sign of the effective potential in matter later.) In what follows the superscript $S$ of $\Phi$ indicates the value of the flux at the production.
\begin{itemize}
\item \textbf{ $\mathcal{V}_\tau>0$ and $\Delta m_{31}^2>0$:}
The ordering of vacuum and effective eigenmasses are $m_1<m_2<m_3$ and  $\tilde{m}_1<\tilde{m}_2<\tilde{m}_3$, respectively.
The adiabacity then implies that the fluxes of neutrinos on the Earth (up to the trivial $r^{-2}$ geometric suppression) are
\begin{equation} \label{PN} \Phi_{\nu_\alpha}^\oplus=\Phi_{\nu_e}^S(\cos^2 \theta |U_{\alpha 1}|^2+\sin^2\theta |U_{\alpha 2}|^2)+\Phi_{\nu_\mu}^S(\sin^2 \theta |U_{\alpha 1}|^2+\cos^2\theta |U_{\alpha 2}|^2) \ .\end{equation}
\item \textbf{ $\mathcal{V}_\tau<0$ and $\Delta m_{31}^2>0$:}
The ordering of vacuum and effective eigenmasses are $m_1<m_2<m_3$ and  $\tilde{m}_3<\tilde{m}_1<\tilde{m}_2$, respectively.
As a result,
\begin{equation} \label{NN}\Phi_{\nu_\alpha}^\oplus=\Phi_{\nu_e}^S(\cos^2 \theta |U_{\alpha 2}|^2+\sin^2\theta |U_{\alpha 3}|^2)+\Phi_{\nu_\mu}^S(\sin^2 \theta |U_{\alpha 2}|^2+\cos^2\theta |U_{\alpha 3}|^2) \ .\end{equation}
\item \textbf{ $\mathcal{V}_\tau>0$ and $\Delta m_{31}^2<0$:}
The ordering of vacuum and effective eigenmasses are $m_3<m_1<m_2$ and  $\tilde{m}_2<\tilde{m}_1<\tilde{m}_3$, respectively.
As a result,
\begin{equation} \label{PI}\Phi_{\nu_\alpha}^\oplus=\Phi_{\nu_e}^S(\cos^2 \theta |U_{\alpha 1}|^2+\sin^2\theta |U_{\alpha 3}|^2)+\Phi_{\nu_\mu}^S(\sin^2 \theta |U_{\alpha 1}|^2+\cos^2\theta |U_{\alpha 3}|^2) \  . \end{equation}
\item \textbf{ $\mathcal{V}_\tau<0$ and $\Delta m_{31}^2<0$:}
The ordering of vacuum and effective eigenmasses are $m_3<m_1<m_2$ and  $\tilde{m}_3<\tilde{m}_2<\tilde{m}_1$, respectively.
As a result,
\begin{equation} \label{NI}\Phi_{\nu_\alpha}^\oplus=\Phi_{\nu_e}^S(\cos^2 \theta |U_{\alpha 2}|^2+\sin^2\theta |U_{\alpha 1}|^2)+\Phi_{\nu_\mu}^S(\sin^2 \theta |U_{\alpha 2}|^2+\cos^2\theta |U_{\alpha 1}|^2) \ .\end{equation}
\end{itemize}
For antineutrinos, we may repeat the same formulas except that we should replace $\mathcal{V}_\alpha$ with $-\mathcal{V}_\alpha$. Considering that IceCube does not distinguish between neutrinos and antineutrinos (see, however, \cite{Skrzypek:2025tmg,Hajjar:2026uzs}), the flavor composition in the Earth will then depend not only on the flavor ratios at the production but also on the ratios of neutrinos and antineutrinos at the source. Let us suppose that at the source, $\pi^+$ and $\pi^-$ are produced with a flavor ratio, $\Phi^S_{\pi^-}/\Phi^S_{\pi^+}=r$. The charged pions decay to muons and neutrinos. Subsequently, muon (anti-muon) decay into a pair of $\nu_\mu$ ($\bar{\nu}_\mu$) and $\bar{\nu}_e$ ($\nu_e$). Then, the fluxes of neutrinos at the source will be $\Phi^S_{\bar{\nu}_e}=r\Phi^S_{{\nu}_e}$ and $\Phi^S_{\nu_\mu}=\Phi^S_{\bar{\nu}_\mu}=\Phi^S_{\bar{\nu}_e}+\Phi^S_{{\nu}_e}$. The value of $r$ is unknown but the values of $r=1$ and $r=3.5$ are theoretically motivated \cite{Skrzypek:2025tmg}.
These predictions for the flavor ratios at Earth combining neutrinos and antineutrinos are shown in  the ternary plot in Fig  \ref{fig:ternery-Vt-Standard-ratio}  taking $r=1$ (thin line)  and $r=3.5$ (thick line) along with the contour plots of the IceCube present results \cite{Abbasi:2025fjc}  and the forecast for IceCube-Gen2 \cite{IceCube-Gen2:2025yte}. Notice that for $r=1$, the initial flux is symmetric under $\nu \leftrightarrow \bar{\nu}$ and therefore under $\mathcal{V}_\alpha \leftrightarrow - \mathcal{V}_\alpha $. As a result, we have only two curves for $r=1$ but four curves for $r=3.5$.  That is the thin red and orange lines are overshadowed respectively by the thin blue and green lines. For normal mass ordering only with $r=3.5$ and $\mathcal{V}_\tau >0$, there will be some chance for IceCube to discriminate between our prediction and the canonic prediction shown with pink color. However, for the inverted mass ordering with $\mathcal{V}_\tau <0$, the predictions for both $r=1$ and $r=3.5$ shown with green lines appear outside the 95 \% C.L. forecast of IceCube-Gen2 which means IceCube-Gen2 can discriminate between our predictions and the canonic ratio. 
This is particularly remarkable, especially that with $U_{e4}=U_{\mu 4}=0$, there will be neither an attenuation of the flux nor an energy spectrum distortion relative to the standard model prediction. Remember that with $\mathcal{V}_e=\mathcal{V}_\mu=0$, there is no  need for the second HNL to be heavier than 500~MeV, permitting light $V$ which can also lead to sizable $\mathcal{V}_\tau$ in even mild spikes with $\gamma_{SP}=3/2$ and $\rho_{spike}=3.2\times 10^{14}$  GeV$/{\rm cm}^3$. Indeed, these results hold valid even without the HNL as long as an alternative model predicts $\mathcal{V}_e=\mathcal{V}_\mu=0$ and $|\mathcal{V}_\tau |>10^{-16}$ eV.  Notice that the predictions are robust against varying $|\mathcal{V}_\tau|$ as long as it is larger than $10^{-16}$ eV.  We have also checked the robustness of our results against varying $\gamma_\nu$ and have found that within $\gamma_\nu \in (2.5,3.2)$, the shifts of  the curves are negligible.  Notice that these results are independent of $\Sigma$ and therefore of  the value of $\gamma_{SP}$ but  depend only on $\rho_{spike} (R_{em})$.

Let us now consider the case $|\mathcal{V}_e|,| \mathcal{V}_\tau|\gg 10^{-17}$ eV  and $\mathcal{V}_\mu=0$. Benchmarks B1 and B2 are representatives of this flavor structure. In this case, flavor eigenstates and effective mass eigenstates coincide in the spike. However, scattering off dark matter can change the flux as described in appendix \ref{cascade}.
Let us denote the fluxes at the exit from spike by $\Phi_{\nu_\alpha}^{SP}$ which can be obtained from the initial flavor ratio using the formulas in appendix \ref{cascade}.  We have solved the differential equation in appendix B to obtain $\Phi_{\nu_\alpha}^{SP}$.
Remember that the signs of $\mathcal{V}_{e}$ and $\mathcal{V}_\tau$ are opposite. Again depending on the signs of $\mathcal{V}_\tau$ and $\Delta m_{31}^2$, four distinct predictions for the neutrino flux ratios at Earth can be made. 

\begin{itemize}
\item \textbf{ $\mathcal{V}_\tau>0$,  $\mathcal{V}_e<0$ and $\Delta m_{31}^2>0$:}
The flavor states at the production point with $-\mathcal{V}_e$ and $\mathcal{V}_\tau>10^{-16}$~eV coincide with the effective mass eigenstates with the ordering from light to heavy as $\nu_e$, $\nu_\mu$ and $\nu_\tau$. They will  end up  in vacuum as $\nu_e\to \nu_1$,
$\nu_\mu \to \nu_2$ and $\nu_\tau \to \nu_3$.
The fluxes on the Earth will be then equal to 
\begin{equation} \label{ePN}\Phi_{\nu_\alpha}^\oplus=\Phi_{\nu_e}^{SP}|U_{\alpha 1}|^2+\Phi_\mu^{SP} |U_{\alpha 2}|^2+ \Phi_{\nu_\tau}^{SP}|U_{\alpha 3}|^2 .\end{equation}
\item \textbf{ $\mathcal{V}_\tau<0$, $\mathcal{V}_e>0$ and $\Delta m_{31}^2>0$:}
The flavor states at the production point with $\mathcal{V}_e$ and $-\mathcal{V}_\tau>10^{-16}$~eV coincide with the effective mass eigenstates with the ordering from light to heavy as $\nu_\tau$, $\nu_\mu$ and $\nu_e$. They will  end up  in vacuum as $\nu_e\to \nu_3$,
$\nu_\mu \to \nu_2$ and $\nu_\tau \to \nu_1$.
The fluxes on the Earth will be then equal to 
\begin{equation} \label{eNN}\Phi_{\nu_\alpha}^\oplus=\Phi_{\nu_e}^{SP} |U_{\alpha 3}|^2+ \Phi_{\nu_\mu}^{SP} |U_{\alpha 2}|^2 +\Phi_{\nu_\tau}^{SP}|U_{\alpha 1}|^2 .\end{equation}
\item \textbf{ $\mathcal{V}_\tau>0$, $\mathcal{V}_e<0$ and $\Delta m_{31}^2<0$:}
The flavor states at the production point with $-\mathcal{V}_e$ and $\mathcal{V}_\tau>10^{-16}$~eV coincide with the effective mass eigenstates with the ordering from light to heavy as $\nu_e$, $\nu_\mu$ and $\nu_\tau$. They will  end up  in vacuum as $\nu_e\to \nu_3$,
$\nu_\mu \to \nu_1$ and $\nu_\tau \to \nu_2$.
The fluxes on the Earth will be then equal to 
\begin{equation} \label{ePI}\Phi_{\nu_\alpha}^\oplus=\Phi_{\nu_e}^{SP} |U_{\alpha 3}|^2+ \Phi_{\nu_\mu}^{SP} |U_{\alpha 1}|^2+ \Phi_{\nu_\tau}^{SP} |U_{\alpha 2}| ^2\ .\end{equation}
\item \textbf{ $\mathcal{V}_\tau<0$, $\mathcal{V}_e>0$ and $\Delta m_{31}^2<0$:}
The flavor states at the production point with $\mathcal{V}_e$ and $-\mathcal{V}_\tau>10^{-16}$~eV coincide with the effective mass eigenstates with the ordering from light to heavy as $\nu_\tau$, $\nu_\mu$ and $\nu_e$. They will  end up  in vacuum as $\nu_e\to \nu_2$,
$\nu_\mu \to \nu_1$ and $\nu_\tau \to \nu_3$.
The fluxes on the Earth will be then equal to 
\begin{equation} \label{eNI}\Phi_{\nu_\alpha}^\oplus=\Phi_{\nu_e}^{SP} |U_{\alpha 2}|^2+\Phi_{\nu_\mu}^{SP} |U_{\alpha 1}|^2+ \Phi_{\nu_\tau}^{SP}  |U_{\alpha 3}|^2 \ .\end{equation}
\end{itemize}
Similar discussion holds valid for antineutrinos with $\mathcal{V}_\alpha \to -\mathcal{V}_\alpha$.

\begin{figure*}[!ht]
	\centering
	\includegraphics[width=1\textwidth ]{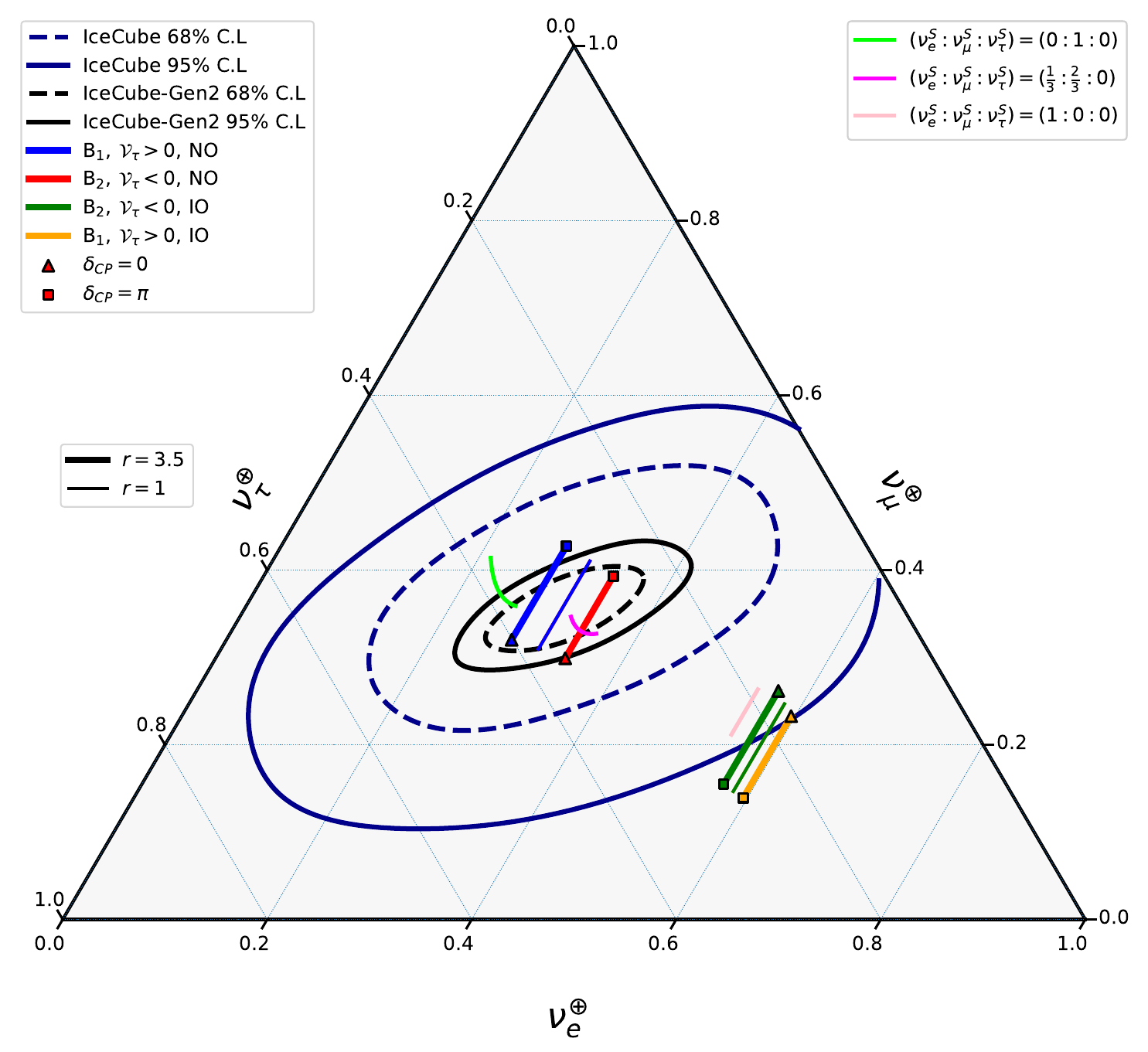}
	\caption{Flavor ratios at the Earth for $\mathcal{V}_\mu=0$, $|\mathcal{V}_e|=9\times10^{-17}\rm eV$  and $|\mathcal{V}_\tau| =1.56\times 10^{-16}\rm eV$,  starting with  $\Phi^S_{\bar{\nu}_e}=r\Phi^S_{\nu_e}$ and  $\Phi^S_{\nu_\mu}=\Phi^S_{\bar{\nu}_\mu}=\Phi^S_{\nu_e}+\Phi^S_{\bar{\nu}_e}$ at the production and varying $0<\delta_{CP}<2\pi$. Thick and thin line segments correspond to $r\equiv\Phi_{\bar\nu_e}^S/\Phi_{\nu_e}^S=3.5$ and $1$, respectively. The dark blue,  red, dark green and orange segments respectively show our predictions for benchmark $\rm B_1$  with  $\Delta m_{31}^2>0$, $\rm B_2$ with $\Delta m_{31}^2>0$, benchmark $\rm B_2$ with  $\Delta m_{31}^2<0$, and benchmark $\rm B_1$ with $\Delta m_{31}^2<0$, respectively.  }
	\label{fig:ternery-Ve-Vt-lowenerg}
\end{figure*}

These predictions are shown in Fig \ref{fig:ternery-Ve-Vt-lowenerg} for benchmarks $\rm B_1$ and $\rm B_2$ for both normal mass ordering and inverted mass ordering, taking $\Phi^S_{\bar{\nu}_e}=r\Phi^S_{\nu_e}$ and  $\Phi^S_{\nu_\mu}=\Phi^S_{\bar{\nu}_\mu}=\Phi^S_{\nu_e}+\Phi^S_{\bar{\nu}_e}$. 
IceCube-Gen2 will not be able to discriminate between the prediction for the flavor ratios with normal ordering and $|\mathcal{V}_e|>10^{-16}$~eV and the standard model prediction but it can distinguish  the prediction for the inverted mass ordering.  With  the inverted neutrino mass ordering and $\pi/2<\delta_{CP}<3 \pi/2$,  the prediction for the $B_1$ and $B_2$ benchmarks  lie outside even the present  2$\sigma$ IceCube flavor contours.  Turning
on $|U_{e 4}|$ leads to the attenuation of the flux (see table \ref{tab:benchmark1}).  The prediction only mildly depends  on the shape of the energy spectrum 
of the neutrino flux  as we vary $\gamma_\nu$ within $(2.5,3.2)$. As seen from the figure the dependence of the prediction on  $r$ is also mild. Like the case of $\mathcal{V}_e=0$ shown in Fig. \ref{fig:ternery-Vt-Standard-ratio},
with $r=1$, the prediction is the same for $\mathcal{V}_\tau>0$ and $\mathcal{V}_\tau<0$.

\begin{figure*}[!ht]
	\centering
	\includegraphics[width=1\textwidth ]{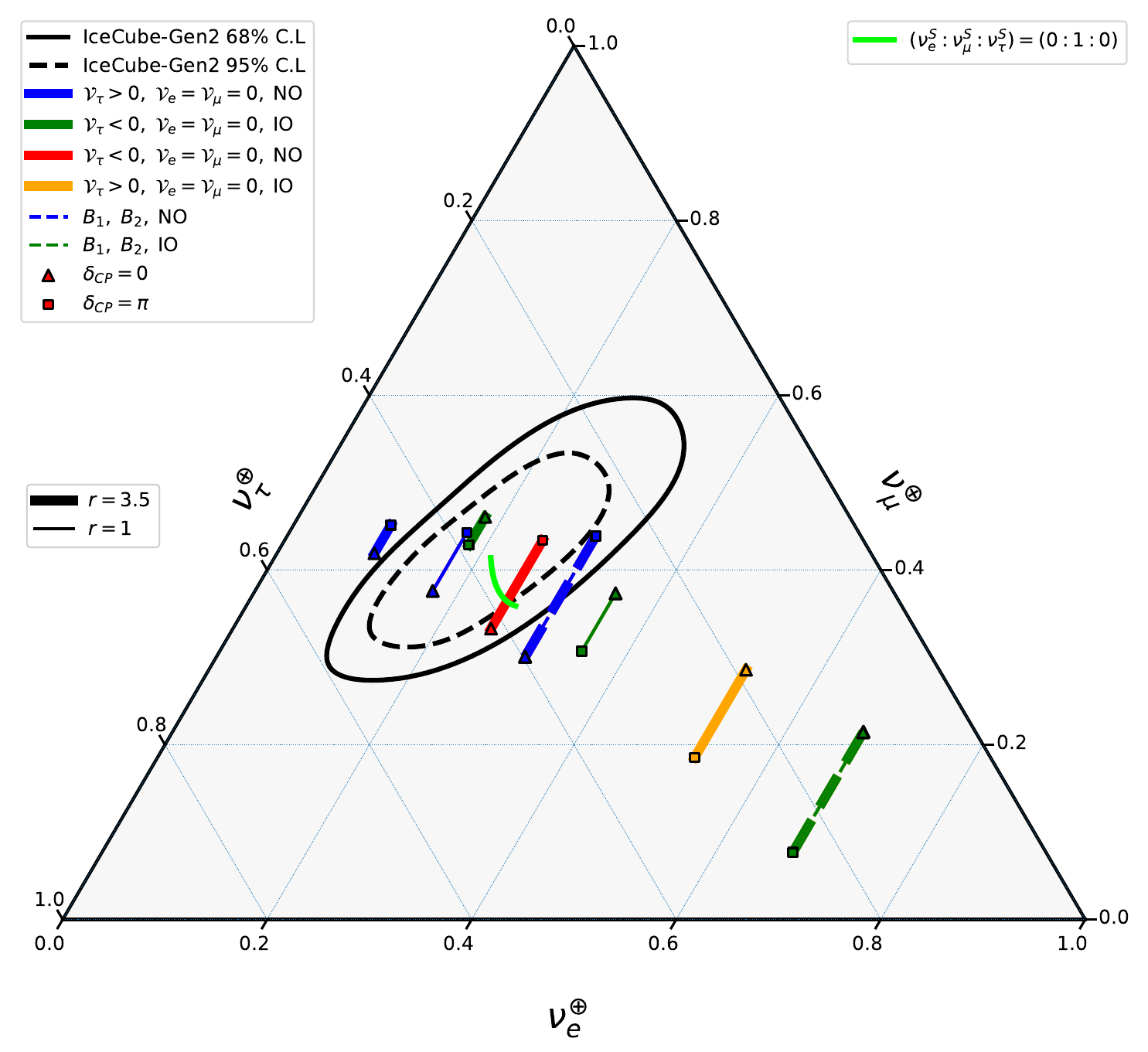}
	\caption{Flavor ratios at the Earth for high energy neutrinos ($E_\nu>$PeV), starting with $\Phi^S_{\nu_e}=\Phi^S_{\bar{\nu}_e}=0$ and  $ \Phi^S_{\bar{\nu}_\mu}=r \Phi^S_{\nu_\mu}$ from a muon damped source and varying $0<\delta_{CP}<2\pi$. The thick and thin line segments correspond to $r=3.5$ and $1$, respectively. 
		The dashed lines correspond to $\mathcal{V}_\mu=0$, $|\mathcal{V}_e|>10^{-17}$ and $|\mathcal{V}_\tau|>10^{-16}$.  The solid lines correspond to   $\mathcal{V}_\mu=\mathcal{V}_e=0$ and $|\mathcal{V}_\tau|>10^{-16}$. The dark blue, dark green, red and orange, lines segments show our predictions respectively for
	$\mathcal{V}_\tau>0$ with $\Delta m_{31}^2>0$, 
	 $\mathcal{V}_\tau<0 $ with $\Delta m_{31}^2<0$,   $\mathcal{V}_\tau<0$ with $\Delta m_{31}^2>0$ and $\mathcal{V}_\tau>0$ with  $\Delta m_{31}^2<0$. The 1$\sigma$ and 2$\sigma$ black contours show the  forecast for the flavor measurement by IceCube-Gen2 under the assumption of a muon damped source within the standard model \cite{IceCube-Gen2:2025yte}. 
	}
	\label{fig:ternaryHIGH}
\end{figure*}

Fig.~\ref{fig:ternaryHIGH} shows the prediction for the flavor ratios at the Earth for high energy neutrinos with $E_\nu>$PeV with $\Phi_{\nu_e}^S=\Phi_{\bar{\nu}_e}^S=0$ and $\Phi_{\bar{\nu}_\mu}^S=r \Phi_{{\nu}_\mu}^S$ at the production from a muon damped source.  For $\mathcal{V}_e=\mathcal{V}_\mu=0$, we have inserted  these initial fluxes  in Eqs. (\ref{PN}-\ref{NI}) to compute the fluxes at the Earth, showing the ratios with solid line segments in Fig.~\ref{fig:ternaryHIGH}. For $\mathcal{V}_\mu=0$ but $\mathcal{V}_e, \mathcal{V}_\tau \ne 0$,  $\nu_\mu$ or $\bar{\nu}_\mu$ produced at the source will leave the spike  without any attenuation. Thus, we can insert $\Phi_{\nu_e}^{SP}=\Phi_{\nu_\tau}^{SP}=0$ but $\Phi_{\bar{\nu}_\mu}^S\ne 0 $ in Eqs.~(\ref{ePN}-\ref{eNI}) to obtain the  neutrino fluxes at the Earth. For anti-neutrinos, we may use the same formulas  changing the sign of $\mathcal{V}_\alpha$. The ratios for this case are shown with dashed  lines. Notice that, in this case, there is no dependence on $\gamma_{\nu}$  and the predictions are also insensitive to the sign of $\mathcal{V}_\tau$ and consequently to that of $\mathcal{V}_e$.  The segments associated to the $\rm B_1$  and $\rm B_2 $ benchmarks therefore overlap with each other for both the normal and inverted mass orderings.  Moreover as expected  for the ${\rm B_1}$  and ${\rm B_2}$ benchmarks,  the thin and thick dashed lines overlap because, the final flavor ratio when summed over the neutrino and antineutrino fluxes should be independent of $r$. The 1$\sigma$ and 2$\sigma$ contour lines show the forecast for IceCube-Gen2 for the standard case with $(\nu_e^S:\nu_\mu^S:\nu_\tau^S)=(0:1:0)$ at the source.  Like  for the low energy neutrinos, IceCube-Gen2 can distinguish the prediction for the inverted mass ordering for the $B_1$ and $B_2$ benchmarks from the standard expectation.  Moreover, with $\mathcal{V}_e=\mathcal{V}_\mu=0$ and $\mathcal{V}_\tau>0$, the flavor ratios for both normal and inverted mass ordering are predicted to stay out of the 95 \% C.L. contour plot forecasts under the standard model.   In the case of normal mass ordering, the discrimination  via  flavor composition of the  neutrino plus antineutrino flux will be more challenging.

Finally, let us discuss the case with $|\mathcal{V}_\mu|,|\mathcal{V}_\tau|\gg 10^{-17}$ eV and $\mathcal{V}_e=0$.  Benchmarks $\rm B_3$ and $\rm B_4$ are representatives of this flavor structure. Again, remembering that the signs of $\mathcal{V}_\mu$ and $\mathcal{V}_\tau$ are opposite, we can distinguish four cases:

\begin{itemize}
	\item \textbf{ $\mathcal{V}_\tau>0$, $\mathcal{V}_\mu<0$ and $\Delta m_{31}^2>0$:}
	The flavor states at the production point with $-\mathcal{V}_\mu$ and $\mathcal{V}_\tau>10^{-16}$~eV coincide with the effective mass eigenstates with the ordering from light to heavy as $\nu_\mu$, $\nu_e$ and $\nu_\tau$. They will  end up  in vacuum as $\nu_e\to \nu_2$,
	$\nu_\mu \to \nu_1$ and $\nu_\tau \to \nu_3$.
	The fluxes on the Earth will be then equal to 
	$$\Phi_{\nu_\alpha}^\oplus=\Phi_{\nu_e}^{SP}|U_{\alpha 2}|^2+\Phi_{\nu_\mu}^{SP} |U_{\alpha 1}|^2+ \Phi_{\nu_\tau}^{SP}|U_{\alpha 3}|^2 .$$
	\item \textbf{ $\mathcal{V}_\tau<0$, $\mathcal{V}_\mu>0$ and $\Delta m_{31}^2>0$:}
	The flavor states at the production point with $\mathcal{V}_\mu$ and $-\mathcal{V}_\tau>10^{-16}$~eV coincide with the effective mass eigenstates with the ordering from light to heavy as $\nu_\tau$, $\nu_e$ and $\nu_\mu$. They will  end up  in vacuum as $\nu_e\to \nu_2$,
	$\nu_\mu \to \nu_3$ and $\nu_\tau \to \nu_1$.
	The fluxes on the Earth will be then equal to 
	$$\Phi_{\nu_\alpha}^\oplus=\Phi_{\nu_e}^{SP} |U_{\alpha 2}|^2+ \Phi_{\nu_\mu}^{SP} |U_{\alpha 3}|^2 +\Phi_{\nu_\tau}^{SP}|U_{\alpha 1}|^2 .$$
	\item \textbf{ $\mathcal{V}_\tau>0$, $\mathcal{V}_\mu<0$ and $\Delta m_{31}^2<0$:}
	The flavor states at the production point with $-\mathcal{V}_\mu$ and $\mathcal{V}_\tau>10^{-16}$~eV coincide with the effective mass eigenstates with the ordering from light to heavy as $\nu_\mu$, $\nu_e$ and $\nu_\tau$. They will  end up  in vacuum as $\nu_e\to \nu_1$,
	$\nu_\mu \to \nu_3$ and $\nu_\tau \to \nu_2$.
	The fluxes on the Earth will be then equal to 
	$$\Phi_{\nu_\alpha}^\oplus=\Phi_{\nu_e}^{SP} |U_{\alpha 1}|^2+ \Phi_{\nu_\mu}^{SP} |U_{\alpha 3}|^2+ \Phi_{\nu_\tau}^{SP}|U_{\alpha 2}|^2 \ .$$
	\item \textbf{ $\mathcal{V}_\tau<0$, $\mathcal{V}_\mu>0$ and $\Delta m_{31}^2<0$:}
	The flavor states at the production point with $\mathcal{V}_\mu$ and $-\mathcal{V}_\tau>10^{-16}$~eV coincide with the effective mass eigenstates with the ordering from light to heavy as $\nu_\tau$, $\nu_e$ and $\nu_\mu$. They will  end up  in vacuum as $\nu_e\to \nu_1$,
	$\nu_\mu \to \nu_2$ and $\nu_\tau \to \nu_3$.
	The fluxes on the Earth will be then equal to 
	$$\Phi_{\nu_\alpha}^\oplus=\Phi_{\nu_e}^{SP} |U_{\alpha 1}|^2+\Phi_{\nu_\mu}^{SP} |U_{\alpha 2}|^2+ \Phi_{\nu_\tau}^{SP}  |U_{\alpha 3}|^2 \ .$$
\end{itemize}

\begin{figure*}[!h]
	\centering
	\includegraphics[width=1\textwidth ]{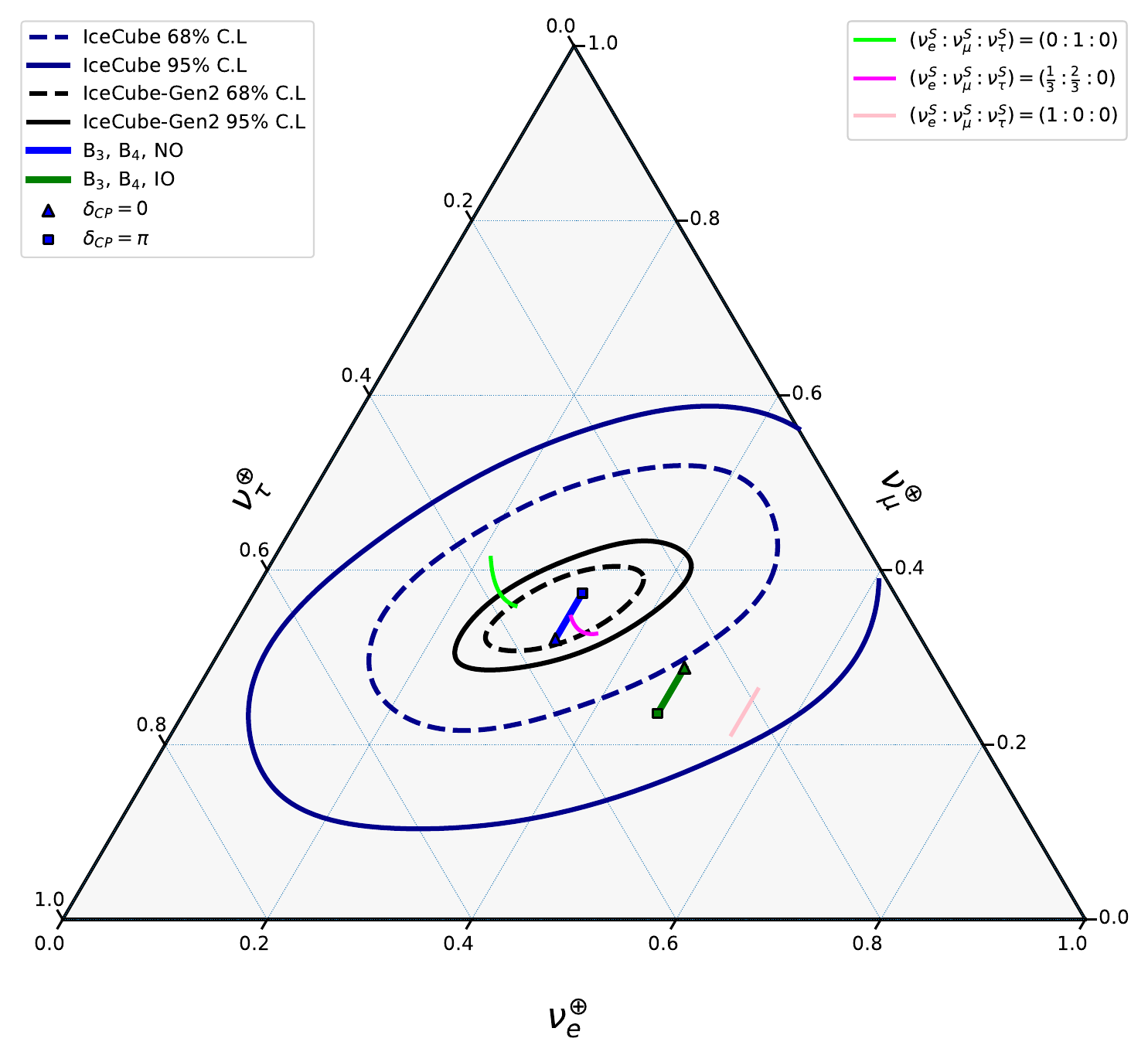}
	\caption{Flavor ratios at the Earth for $\mathcal{V}_e=0$, $|\mathcal{V}_\mu|=1.2\times10^{-16}\rm eV$  and $|\mathcal{V}_\tau| =1.2\times 10^{-14}\rm eV$,  starting with  $\Phi^S_{\bar{\nu}_e}=r\Phi^S_{\nu_e}$ and  $\Phi^S_{\nu_\mu}=\Phi^S_{\bar{\nu}_\mu}=\Phi^S_{\nu_e}+\Phi^S_{\bar{\nu}_e}$ at the production and varying $0<\delta_{CP}<2\pi$. The dark blue and dark green segments respectively show our predictions for benchmark $B_3$ and $B_4$ with $\Delta m_{31}^2>0$ and benchmark $B_3$ and $B_4$ with $\Delta m_{31}^2<0$.  The results are independent of both the sign of  $\mathcal{V}_\tau$ and  the value of $r=\Phi^S_{\bar{\nu}_e}/\Phi^S_{\nu_e}$.  }
	\label{fig:ternery-Vmu-Vt-lowenerg}
\end{figure*}

Like the $B_1$ and $B_2$ benchmarks, the fluxes of the antineutrinos at the Earth are given by the same formulas, replacing $\mathcal{V}_\alpha$ with $-\mathcal{V}_\alpha$. At the $\rm B_3$ and $\rm B_4$ benchmarks, the fluxes of $\nu_e$ and $\bar{\nu}_e$ do not undergo attenuation so 
 $\Phi_{\bar{\nu}_e}^{SP}=	\Phi_{\bar{\nu}_e}^{S}=r \Phi_{{\nu}_e}^{SP}=r \Phi_{{\nu}_e}^{S}$. However, the fluxes of $ \Phi_{\nu_\mu}$ and $\Phi_{\bar{\nu}_\mu}$ undergo equal attenuation and in the process produce equal  fluxes of $\nu_\tau$ and $\bar{\nu}_\tau$. 
 That is $\Phi_{\bar{\nu}_\mu}^{SP}=\Phi_{{\nu}_\mu}^{SP}=s_1 (\Phi_{{\nu}_e}^{S}+	\Phi_{\bar{\nu}_e}^{S})$ and  $\Phi_{\bar{\nu}_\tau}^{SP}=\Phi_{{\nu}_\tau}^{SP}=(s-s_1) (\Phi_{{\nu}_e}^{S}+	\Phi_{\bar{\nu}_e}^{S})$ such that $0<s_1<s<1$. We have obtained the values of $\Phi_{\nu_\alpha}^{SP}$ and $\Phi_{\bar\nu_\alpha}^{SP}$  in terms of $\Phi_{\nu_\alpha}^{S}$ and $\Phi_{\bar\nu_\alpha}^{S}$ (or equivalently the values of $s$ and $s_1$) by solving the differential equation in appendix B.
  Then, regardless of the sign of $\mathcal{V}_\tau$,  the sum of the fluxes of $\nu_\alpha $ and $\bar\nu_\alpha $ reaching the Earth will be
 $$ (\Phi_{\nu_e}^{S}+ \Phi_{\bar{\nu}_e}^{S}) (|U_{\alpha 2}|^2+s(|U_{\alpha 1}|^2+|U_{\alpha 3}|^2))$$
 for normal mass ordering and
  $$ (\Phi_{\nu_e}^{S}+ \Phi_{\bar{\nu}_e}^{S}) (|U_{\alpha 1}|^2+s(|U_{\alpha 2}|^2+|U_{\alpha 3}|^2))$$
  for inverted mass ordering.  If there was no attenuation of the total flux, $s$ would be equal to one and then the ratios would be independent of the neutrino mixing parameters because $|U_{\alpha 1}|^2+|U_{\alpha 2}|^2+|U_{\alpha 3}|^2=1$. However, from tab. \ref{tab:benchmark1}, we see that there should be non-negligible attenuation so $s<1$. Notice that the flavor ratios turn out not only independent of the sign of $\mathcal{V}_\alpha$ but also independent of the ratio $r= \Phi^S_{\bar{\nu}_e}/ \Phi^S_{{\nu}_e}$. 
These predictions are shown in Fig \ref{fig:ternery-Vmu-Vt-lowenerg} for the $\rm B_3$ and $\rm B_4$ benchmarks for both normal and inverted mass ordering. IceCube-Gen2  with this method will be able to distinguish between the standard model prediction and our prediction in case of inverted mass ordering but not for the case of normal mass ordering.

By the time that IceCube-Gen2 collects enough data for the neutrino flavor measurement, the terrestrial neutrino experiments, JUNO and long baseline neutrino experiments are expected to determine the neutrino mass ordering as well as the value of $\delta_{CP}$.  Moreover, $V$ and $N_\tau$ may be discovered at the low energy experiments studying the $\tau$ decay modes or other setups.  Similarly, hints for $N_\mu$ and/or $N_e$  may appear as the violation of the unitarity of the $3 \times 3$ PMNS matrix. The flavor measurement by IceCube-Gen2 then provides information about the spikes and dark matter composing it. To be specific, it provides information on $(g_\psi/m_\psi)[\rho_{spike}(R_{em})-\bar{\rho}_{spike}(R_{em})]$  in which $\rho_{spike}(R_{em})$ and $\bar{\rho}_{spike}(R_{em})$ are respectively the densities of $\psi$ and $\bar{\psi}$.

The following remarks are in order:
(1)
If the bounds on $|U_{e 4}|^2$ and $|U_{\mu 4}|^2$ are tightened by a factor larger than 3 but $V$ and $N_\tau$ are discovered, according to Fig. \ref{fig:ternery-Vt-Standard-ratio}, we expect  a large deviation from the standard prediction for inverted mass ordering with  $\mathcal{V}_\tau <0$. If inverted mass ordering is established but such a deviation is not observed, we can either  extract an upper bound on 
$(g_\psi/m_\psi)[\rho_{spike}(R_{em})-\bar{\rho}_{spike}(R_{em})]$ or conclude that $g_\psi g_N>0$.  The establishment  of such a  deviation can be considered as a great hint for asymmetric sub-GeV dark matter with $g_\psi g_N<0$. (2) Let us suppose $V$, $N_\tau$ and $N_e$ ($N_\mu$) with masses in the range $(50~{\rm MeV},{\rm GeV})$ and with couplings close to the present upper bound are discovered. If JUNO or other experiments establish normal mass ordering, finding a flavor ratio close to $(1/3,1/3,1/3)$ for $E_\nu<1$ PeV by IceCube-Gen2 cannot rule out this scenario. However, if inverted mass ordering is established, we will expect a deviation towards $\nu_e^\oplus$ larger than $1/3$.  A flavor ratio close to $(1/3,1/3,1/3)$ for $E_\nu<1$ PeV can then be interpreted as an upper bound on $(g_\psi/m_\psi)[\rho_{spike}(R_{em})-\bar{\rho}_{spike}(R_{em})]$.

 Notice that the sensitivities of our results  to quantities such as  $R_{em}$  are only through $\mathcal{V}_\alpha$ and the column density, $\Sigma$. As emphasized, for $|\mathcal{V}_\alpha|>10^{-16} $ eV, the results are stable against varying, $\mathcal{V}_\alpha$.   With $g_\psi g_N|U_{\tau 5}|^2/m_V^2\sim 5\times 10^{-6}$~MeV$^{-2}$ and $m_\psi\sim 0.1$ GeV, we can have $|\mathcal{V}_\tau|>10^{-16}$ eV  for $R_{em}<700~R_S$.  Thus, our results for the flavor configuration $\mathcal{V}_e=\mathcal{V}_\mu=0$ and $\mathcal{V}_\tau \ne 0$ is robust against the value of $R_{em}$. However, considering the strong bounds on $|U_{\mu 4}|^2$ and $|U_{e 4}|^2$, the conditions $|\mathcal{V}_\mu|>10^{-16}$ eV and/or   $|\mathcal{V}_e|>10^{-16}$ eV cannot be fulfilled if $R_{em}\stackrel{>}{\sim } 50 R_S$ so the predictions for the flavor configurations in which $\mathcal{V}_\mu$ or $\mathcal{V}_e$ are less robust  against varying $R_{em}$.  As emphasized before, our analysis is based on the favorable case of the NGC 1068 spike with $\gamma_{SP}=7/3$ which is predicted to have largest $\rho_{spike}(R_{em})$ among the four AGN sources identified so far (see tab.~\ref{table:parameters}). 
 For lower density spikes, the effect discussed in the present paper will be too small. However, we should remember that NGC 1068 is a steady powerful source of high energy neutrinos. Based on the IceCube data collected during 2011-2020, eighty events have already been discovered originated by NGC 1068 \cite{IceCube:2022der}.
 We therefore expect substantial number of events from this point source by the time that IceCube-Gen2 comes to the end of its mission. Thus, even this single source can teach us about the potential new physics effect discussed in the present paper.

\begin{figure*}
	\centering
	
	\subfloat[]{%
		\parbox{0.46\textwidth}{%
			\begin{overpic}[width=\linewidth]{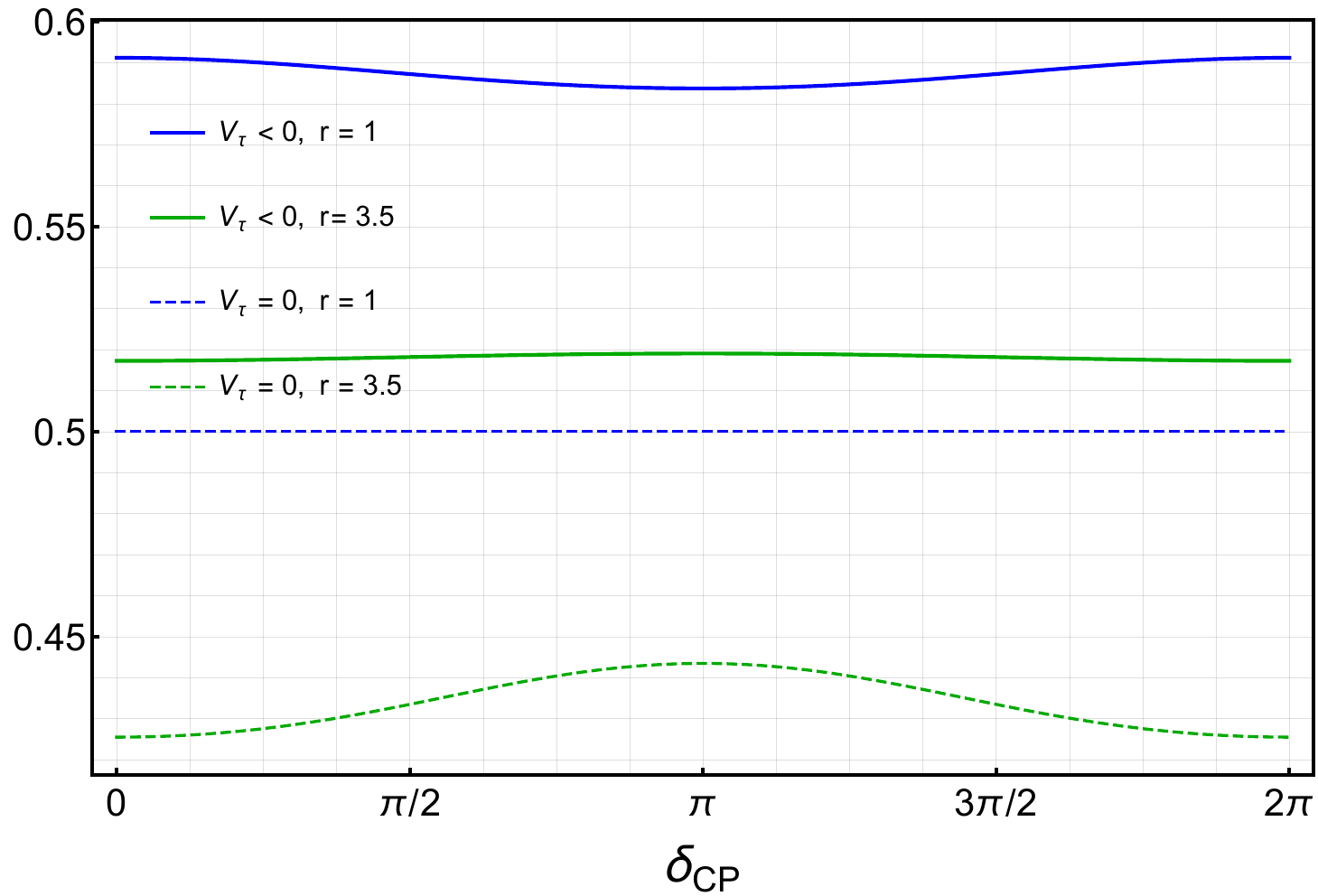}
				\put(-5,20){\rotatebox{90}{%
						$\displaystyle
						\Phi_{\nu_\mu}^\oplus/
						(\Phi_{\nu_\mu}^\oplus+\Phi_{\bar{\nu}_\mu}^\oplus)$
				}}
			\end{overpic}%
		}%
		\label{fig:Mu-OnlyVta-Negative}%
	}
	\hfill
	\subfloat[]{%
		\parbox{0.46\textwidth}{%
			\begin{overpic}[width=\linewidth]{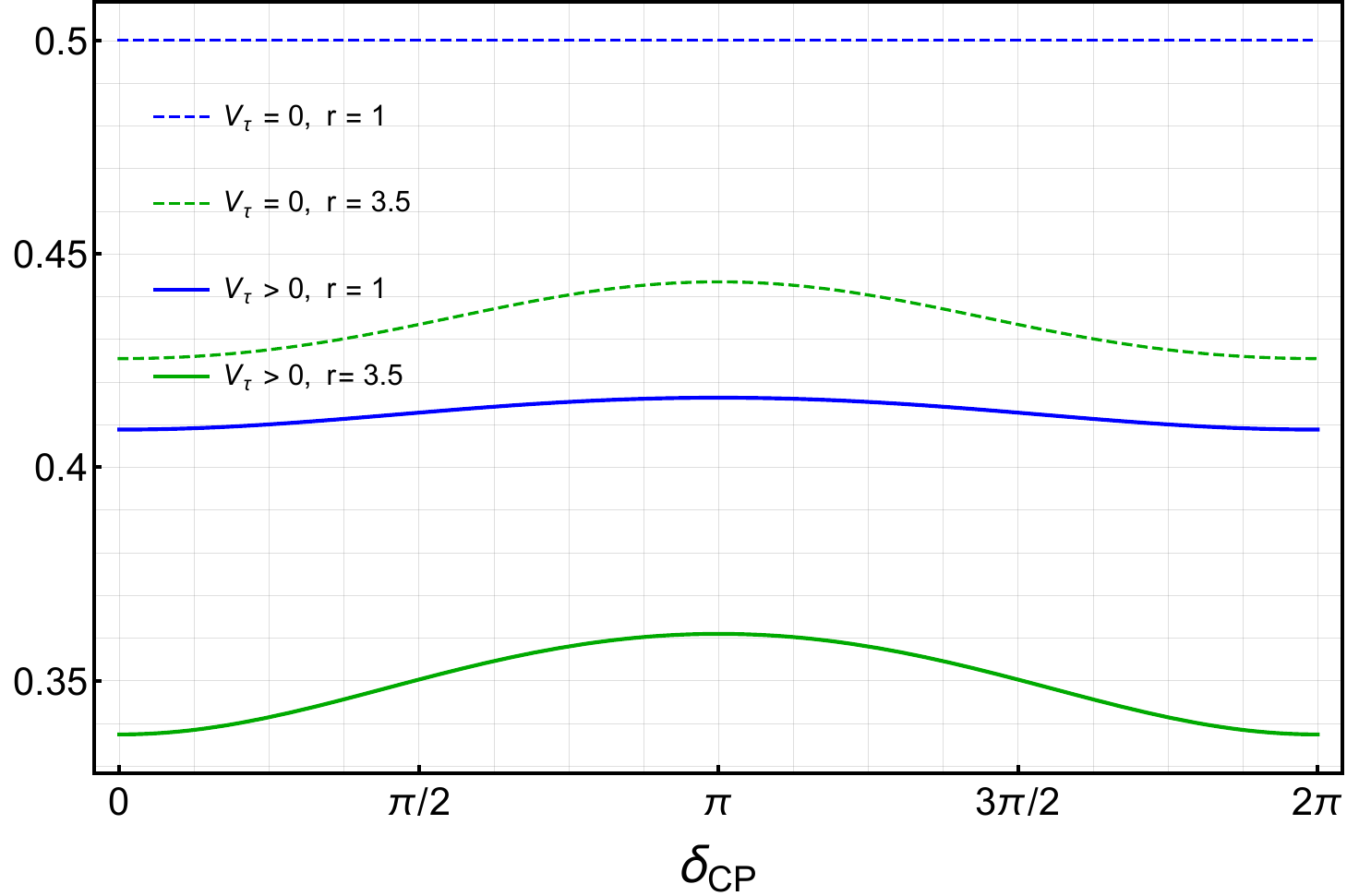}
				\put(-5,20){\rotatebox{90}{%
						$\displaystyle
						\Phi_{\nu_\mu}^\oplus/
						(\Phi_{\nu_\mu}^\oplus+\Phi_{\bar{\nu}_\mu}^\oplus)$
				}}
			\end{overpic}%
		}%
		\label{fig:Mu-OnlyVta-positive}%
	}
	
	\vspace{0.5cm}
	
	\subfloat[]{%
		\parbox{0.46\textwidth}{%
			\begin{overpic}[width=\linewidth]{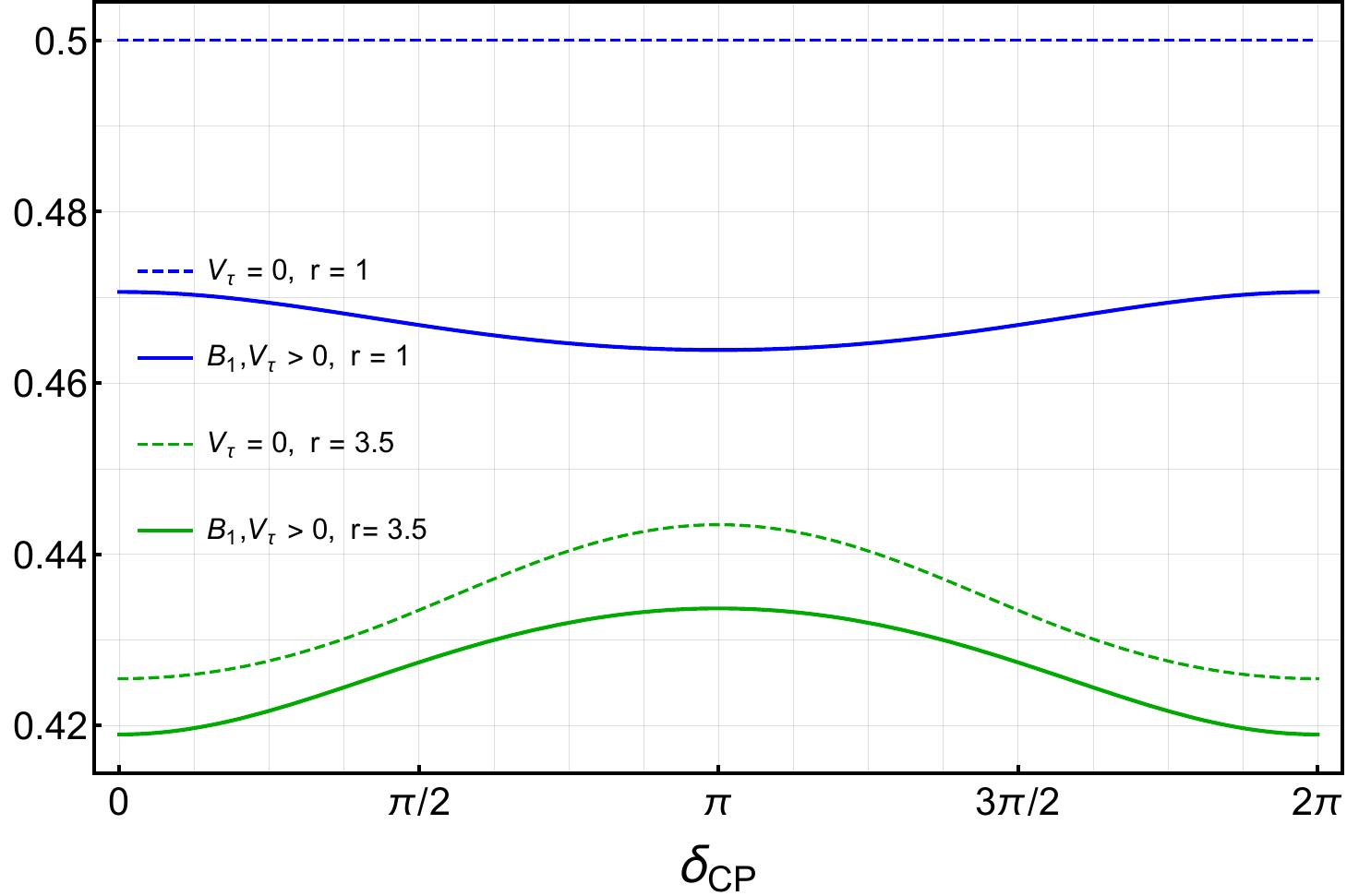}
				\put(-6,20){\rotatebox{90}{%
						$\displaystyle
						\Phi_{\nu_\mu}^\oplus/
						(\Phi_{\nu_\mu}^\oplus+\Phi_{\bar{\nu}_\mu}^\oplus)$
				}}
			\end{overpic}%
		}%
		\label{fig:Mu-B1N}%
	}
	\hfill
	\subfloat[]{%
		\parbox{0.46\textwidth}{%
			\begin{overpic}[width=\linewidth]{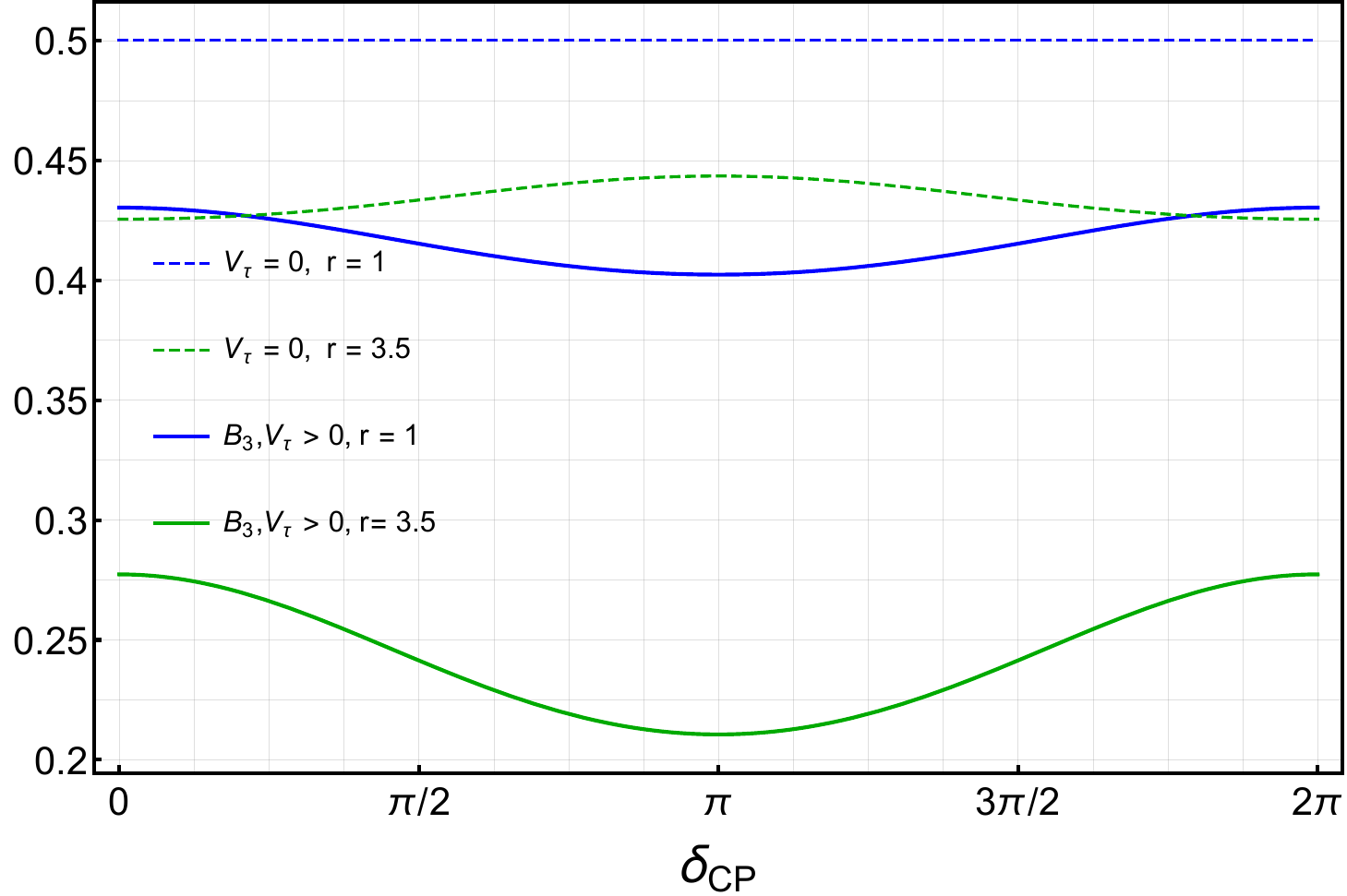}
				\put(-6,20){\rotatebox{90}{%
						$\displaystyle
						\Phi_{\nu_\mu}^\oplus/
						(\Phi_{\nu_\mu}^\oplus+\Phi_{\bar{\nu}_\mu}^\oplus)$
				}}
			\end{overpic}%
		}%
		\label{fig:Mu-B3N}%
	}
	
	\caption{Muon-neutrino flux fraction at Earth, $\Phi_{\nu_\mu}^\oplus/\left(\Phi_{\nu_\mu}^\oplus+\Phi_{\bar{\nu}_\mu}^\oplus\right)$ versus  $\delta_{CP}$ for normal mass ordering $(\Delta m_{31}^2>0)$, starting with $\Phi_{\bar{\nu}_e}^S=r\Phi_{\nu_e}^S$ and  $\Phi_{\nu_\mu}^S=\Phi_{\bar{\nu}_\mu}^S=\Phi_{\nu_e}^S+\Phi_{\bar{\nu}_e}^S$ at the source. The dark-blue and dark-green curves correspond to $r\equiv\Phi_{\bar{\nu}_e}^S/\Phi_{\nu_e}^S=1$ and $r=3.5$, respectively.  The solid curves represent the predictions including the non-standard interaction. The dashed curves show the standard prediction with  with $\mathcal{V}_\alpha=0$. The upper panels show the predictions for $\mathcal{V}_e=\mathcal{V}_\mu=0$. Panels (a) and (b) respectively correspond to $\mathcal{V}_\tau<0$ and $\mathcal{V}_\tau>0$. Panels (c)  and (d) show the results for the benchmarks $B_1$ and $B_3$, both with $\mathcal{V}_\tau>0$. }
	\label{fig:Muon-over-muAntimu}
\end{figure*}

 From Figs. (\ref{fig:ternery-Vt-Standard-ratio},\ref{fig:ternery-Ve-Vt-lowenerg},\ref{fig:ternaryHIGH},\ref{fig:ternery-Vmu-Vt-lowenerg}), we observe that sizeable deviations from the standard prediction are possible mainly for the inverted mass ordering. For normal mass ordering, the deviations from the standard flavor ratio prediction for neutrino and antineutrino components cancel each other. If IceCube can differentiate between the neutrino and antineutrino fluxes, the chances to discern the new physics effects predicted by our scenario will increase.  As is well-known, the inelasticity distribution of the deep inelastic scatterings of neutrinos and antineutrinos are distinct from each other. With increased statistics, discrimination between the fluxes of $\nu_\mu$ and $\bar{\nu}_\mu$ can be possible \cite{Skrzypek:2025tmg}; see also \cite{Hajjar:2026uzs}. Bearing in mind these developments \cite{Skrzypek:2025tmg,Hajjar:2026uzs}, let us study $\Phi_{\nu_\mu}^\oplus/(\Phi_{\nu_\mu}^\oplus+ \Phi_{\bar\nu_\mu}^\oplus)$ for the normal mass ordering. The dependence  of this ratio on $\delta_{CP}$ is shown in Fig. \ref{fig:Muon-over-muAntimu} for normal mass ordering for different flavor structures of $\mathcal{V}_\alpha$. For the two cases of $\mathcal{V}_e= \mathcal{V}_\mu=0$ and $\mathcal{V}_\tau>0$ (shown in Fig. \ref{fig:Muon-over-muAntimu}-b) and benchmark $B_3$ (shown in Fig. \ref{fig:Muon-over-muAntimu}-d), the deviation due to nonzero $\mathcal{V}_\tau$ (the separation of the solid and dashed lines) is significant.  The case of $B_3$ is of particular  interest  because, as shown in Fig. \ref{fig:ternery-Vmu-Vt-lowenerg}, when the fluxes of neutrinos and antineutrinos  are combined, the sensitivity to $\mathcal{V}_\tau$ is completely lost for the normal mass ordering. Fig.~\ref{fig:Muon-over-muAntimu}  demonstrates that the measurement of  $\Phi_{\nu_\mu}^\oplus/(\Phi_{\nu_\mu}^\oplus+ \Phi_{\bar\nu_\mu}^\oplus)$ will be a powerful tool not only to learn about $r=\Phi_{\bar{\nu}_e}^S/\Phi_{{\nu}_e}^S$ but also about the new physics effects. In particular, we predict a significant deviation between the case  of $r=3.5$ with  $\mathcal{V}_\tau>0$ and that of  $r=1$ with  $\mathcal{V}_\tau=0$.

\section{Conclusions and Discussion \label{con}}

We have built a $U_{NEW}(1)$ gauge model for the interaction of neutrinos with the dark matter particles. The dark matter in our mode is a Dirac fermion, $\psi$ charged under  $U_{NEW}(1)$. The gauge symmetry is spontaneously broken by a new light scalar, $\phi$, charged under $U_{NEW}(1)$. In addition to $\phi$, $V$ and $\psi$, the model contains two heavy neutral leptons (HNLs) whose left-handed components have opposite charges under $U_{NEW}(1)$  to cancel the $[U_{NEW}(1)]^3$  anomaly. After gauge symmetry breaking, one of these HNLs ($N_\tau$)
mixes with $\nu_\tau$.  Thanks to this mixing, $\nu_\tau$ can obtain a coupling of form $V_\mu\bar{\nu}_\tau \gamma^\mu N_\tau $ as well as of form $V_\mu\bar{\nu}_\tau \gamma^\mu \nu_\tau $. The former leads to
 $\nu_\tau+\psi \to N_\tau+\psi$ followed by $N_\tau\to \nu_\tau \bar{\nu} \nu$. The mediator is a relatively light gauge boson so in the scatterings with  the center of mass energy much higher that the mass mediator, the total scattering cross section  can still be sizable. Moreover, the final neutrinos
will carry approximately all the energy of the initial neutrino.
In our model, the new particles have masses less than a few 100 MeV.  
The other HNL  may 
mix with $\nu_e$ or $\nu_\mu$, leading to similar scattering off $\psi$ in the spike.

We have assumed the dark matter production mechanism in the early universe to be asymmetric  \cite{Nussinov:1985xr,Barr:1990ca,Barr:1991qn,Kaplan:1991ah,Davoudiasl:2012uw,Zurek:2013wia, Dev:2026ddq} so that the spike remains immune against annihilation. We have also studied coannihilation dark matter freeze-out mechanism as an alternative but in this case, there will be an unavoidable $p$-wave  annihilation channel to $\nu_\tau\bar{\nu}_\tau  $  at one loop level with a cross section large enough to lead to the destruction of the inner parts of the spike.

In our model,   $N_\tau$,  has a  relatively large mixing with $\nu_\tau$, $|U_{\tau 5}|^2\sim 0.01$. Since the dominant decay mode of $N_\tau$ is into neutrinos, it can avoid the strong bounds on the mixing from the  $N_\tau$ searches, looking for the charged decay products of  $N_\tau$ in various terrestrial experiments. However, $N_\tau$ and $V$ can be discovered by  improved measurement of various $\tau$ decay modes and the energy distribution of the $\tau$ decay products. Moreover, the light gauge boson can be discovered by improving the measurements of the invisible decay mode of the standard model $Z$ gauge boson.
Thus, both the new gauge boson and $N_\tau$ are around the corner as their couplings are close to the present bounds. 

The other HNL can have a mixing with $\nu_e$ or $\nu_\mu$ of order of $|U_{e 4}|^2\sim 10^{-3}$ or $|U_{\mu 4}|^2\sim 10^{-4}$ provided that the second HNL is heavier than the Kaon mass. Since the masses of both HNLs and $V$ are given by the same $\langle \phi \rangle$, this implies $m_V/g_N \gtrsim 250$ MeV in the perturbative regime.
With $U_{e 4}=U_{\mu 4}=0$, this lower bound on $m_V$ can be relaxed implying that  the $\nu_\tau$ interaction with $\psi$ can be significant even in the spikes whose inner densities have been reduced by stellar heating or other mechanisms ({\it i.e.,} spikes with mild inner density described with $\gamma_{SP}\simeq 3/2$.) 
If the mixing of $\nu_e$ or $\nu_\mu$ is large, the footprint of the second HNL can show up in the unitarity violation of the first or second row of the $3\times 3$ PMNS mixing matrix.  Notice that in our model lepton flavor is conserved. Moreover, at the tree level, quarks or charged leptons do not couple to $V$.

Since the dark matter density is taken to be asymmetric,  neutrinos coupled to $V$ obtain an effective potential at the production point in the spike because of the forward scattering off $\psi$ in the dense region. We have focused on $|\mathcal{V}_\tau|>10^{-16}$~eV so that $\mathcal{V}_\tau$ will dominate over $\Delta m_{31}^2/(2 E_\nu)$ in the Hamiltonian for $E_\nu >1$~PeV  while $\nu_\tau$ and $\bar{\nu}_\tau$ propagate in the inner parts of the spike.  We have shown that the evolution of the neutrino flavor in the spike is adiabatic. We have studied the impact of the spike on the flavor composition of the flux reaching the Earth, taking into account both the adiabatic evolution in the $\psi$ background and the inelastic scattering, $\nu+\psi \to N+\psi$ followed by $N\to \nu \bar{\nu}\nu$. We have discussed three distinct regimes with a robust prediction for the final flavor ratios.

(i) $|\mathcal{V}_\tau|>10^{-16}$~eV but $\mathcal{V}_e=\mathcal{V}_\mu=0$: In this case, $\nu_e$ and $\nu_\mu$ cannot oscillate to $\nu_\tau$. Considering that in this case only $\nu_\tau$ couple to $V$, the flux will not  suffer from attenuation. However, the adiabatic evolution may lead to large deviation of the final flavor ratios from the standard model prediction which can be discernible by IceCube-Gen2. As shown in Fig.~\ref{fig:ternery-Vt-Standard-ratio},  the final ratios for $E_\nu<$PeV depend on the mass ordering as well as  on the sign of $\mathcal{V}_\tau$, determined by the relative sign of the gauge couplings of $\psi$ and $N_\tau$,  ${\rm sgn}(g_\psi g_N)$. For inverted  mass ordering and $\mathcal{V}_\tau<0$, the deviation is predicted to be large enough for IceCube-Gen2 to  resolve. For normal mass ordering, the deviations from the standard prediction for neutrino and antineutrino fluxes are opposite each other and cancel out. However, by  the newly-developed methods of separately measuring the $\nu_\mu$ and $\bar{\nu}_\mu$ fluxes \cite{Skrzypek:2025tmg,Hajjar:2026uzs}, detecting the deviation may become  possible for normal mass ordering as demonstrated in Fig. \ref{fig:Muon-over-muAntimu}. 
The final ratios for $E_\nu>$PeV  from  a muon damped source are shown in Fig.~\ref{fig:ternaryHIGH}.  Again, for inverted mass ordering, IceCube-Gen2 will have a better chance to resolve the deviation from the standard prediction.
Notice that these predictions are robust against varying the spike profile within a reasonable range $(3/2<\gamma_{SP}<7/3$) and varying $|U_{\tau 5}|^2$ in $(10^{-4},10^{-2})$.  In this scenario, we expect  $N_\tau$ as well as $V$
to show up in the terrestrial experiments such as the setups studying the $\tau$ decay mode. Then,  the measurement of the flavor ratio provides information on the spike and the dark matter inside it through the combination $(g_\psi/m_\psi)[\rho_{spike}(R_{em})- \bar{\rho}_{spike}(R_{em})]$.

(ii) $|\mathcal{V}_e|,|\mathcal{V}_\tau|>10^{-16}$ eV: In this case,  inelastic scattering and therefore attenuation can take place in the spike but the total flux suppression cannot exceed $1/3$ meaning that more than $2/3$ of the total flux will survive.  Moreover the power law behavior of the energy spectrum will be maintained with the same value of the spectral index. The flux flavor composition can however dramatically deviate  from the standard prediction detectable by IceCube-Gen2 as shown in Figs.~(\ref{fig:ternery-Ve-Vt-lowenerg},\ref{fig:ternaryHIGH}).  Even present flavor ratio measurements by IceCube can rule out certain ranges of the parameters as demonstrated in  Fig.~\ref{fig:ternery-Ve-Vt-lowenerg}. For this scenario we expect $N_e$ as well as $N_\tau$ and $V$ to be discovered by future improvements in the unitarity tests of the PMNS matrix  and the tau decay mode measurements. Remembering that the neutrino mass ordering and $\delta_{CP}$ will be determined by  the upcoming neutrino experiments,  the flavor measurement by IceCube-Gen2 can   then provide information on  $(g_\psi/m_\psi)[\rho_{spike}(R_{em})- \bar{\rho}_{spike}(R_{em})]$.

(iii) $|\mathcal{V}_\mu |,|\mathcal{V}_\tau|>10^{-16}$:  Similarly to the previous case,  adiabatic flavor conversion and inelastic scattering can  simultaneously  affect the flavor composition of the flux traversing the spike, leading to a deviation of the flavor ratio from the standard model prediction. In the case of inverted mass ordering, regardless of the sign of $\mathcal{V}_\alpha$ and  independently of the exact value of the power of the neutrino energy spectrum, $\gamma_\nu$,  the deviation of  flavor ratios of neutrino plus antineutrino fluxes  from the standard predictions will be    discernible by IceCube-Gen-2  as shown in Fig.~\ref{fig:ternery-Vmu-Vt-lowenerg}. However, in the case of normal mass ordering the deviation for neutrinos and antineutrinos cancel each other, but like the case (i),  separate measurement of the fluxes of $\nu_\mu$ and $\bar{\nu}_\mu$ \cite{Skrzypek:2025tmg,Hajjar:2026uzs} can reveal the effect of new physics as shown in Fig. \ref{fig:Muon-over-muAntimu}-d. The attenuation of the flux can be down to 1/3 which can still be tolerated within the uncertainties. The form of energy spectrum will  however be maintained. In this case, in addition to $N_\tau$ and $V$, $N_\mu$ can be discovered by terrestrial experiments testing the unitarity of the PMNS matrix. The flavor measurement then can provide information on $(g_\psi/m_\psi)[\rho_{spike}(R_{em})- \bar{\rho}_{spike}(R_{em})]$.

Our results demonstrate that the flavor ratio measurements combined with the measurement of $\Phi_{\nu_\mu}^\oplus/(\Phi_{\nu_\mu}^\oplus+\Phi_{\bar\nu_\mu}^\oplus)$ \cite{Skrzypek:2025tmg,Hajjar:2026uzs}  can be a powerful tool for searching for new physics even in the absence of spectral distortion or a detectable total flux reduction. 
\appendix
\section{Depletion of the dark matter spike by the neutrino flux }\label{depletion}

In the following, we evaluate whether the neutrino flux from NGC 1068 can destroy the spike.
We take the distance from NGC 1068 to the Earth, $D_L$, equal to 14.4 Mpc. For the neutrino flux at the Earth, we take  \cite{IceCube:2022der}
$$\Phi_{\nu_e+\bar{\nu}_e}^\oplus = \Phi_{\nu_\mu+\bar{\nu}_\mu}^\oplus=\Phi_{\nu_\tau+\bar{\nu}_\tau}^\oplus=1.5\times 10^{-10} \left( \frac{{\rm TeV}}{E_\nu}\right)^{3.2}  {\rm TeV}^{-1}{\rm cm}^{-2}{\rm sec}^{-1}\ . $$
This is the value  inferred from the IceCube data in Ref. \cite{IceCube:2022der} for $E_\nu \in [1.5~{\rm TeV}, 15~{\rm TeV}]$. In the absence of the neutrino attenuation en route, the flux at source would be  $$\Phi^s_{\nu+\bar{\nu}}=\Phi^\oplus_{\nu+\bar{\nu}} \frac{D_L^2}{r^2}$$ where $r$ is the distance from the production. Taking $\Sigma \sigma/m_\psi \sim 1$.  The attenuation would imply a factor of $O(1)$ larger flux  at the source but this would not alter the overall picture. Let us evaluate the required duration  of the flux for all $\psi$ in the spike on the way of the neutrinos to be scattered
$$ \Delta t \sim  \frac{\int \int_{R_{min}}^{R_{max}}(\rho_{spike}/m_\psi) r^2 d\Omega dr}{\int_{R_{min}}^{R_{max}} \int (\rho_{spike}/m_\psi) r^2 \int_{E_{min}}^{E_{max}} \sigma(E_\nu) \Phi_{\nu_\tau +\bar{\nu}_\tau}(E_\nu) dE_\nu  d\Omega dr}\sim 10^{10}~{\rm years} \frac{10^{-30} cm^{2}}{\sigma(E_\nu)} \ ,$$
where we have taken $\rho_{spike}=\rho_0 (r_0/r)^{\gamma_{SP}}$ with $\gamma_{SP}=7/3$, $R_{min}=3 \times 10^{-4}$ pc, $R_{max}=7 R_{min}$, $E_{min}=0.1$ TeV and $E_{max}=15$~TeV. Notice that we have taken a cross section constant in energy with a value that at $m_{\psi}=0.05$ GeV leads to $\sigma \Sigma/m_{\psi} \sim 1$. The required duration is much larger than  the spike reassembling time scale  which is  $10^8$ years. Thus, our conclusion is that the intensity of the neutrino flux will not be strong  enough to destroy the spike,  even if the cross section is so large that  each neutrino undergoes scattering.
Increasing $R_{max}$ from $7R_{min}$ to 700 pc (the size of the spike),  the required duration even exceeds $10^{14}$ years, further confirming this conclusion.

In the above discussion, we showed that the scattering of the high energy neutrino flux  off the spike of NGC 1068 cannot destroy it.  However, as discussed in \cite{Herrera:2026hal,Bertone:2024rxe},  other phenomena such as stellar heating can reduce the density of the spike. Exploring all these details is beyond the scope of the present paper.

\section{Cascade equations }\label{cascade}
In this appendix, we first compute the spectrum of $\nu$ from the chain decay, $N\to \nu V$ followed by $V\to \nu\bar{\nu}$. We then write down the cascade equation for the two neutrino case.
The energy spectrum  of $\nu$ from $N\to V\nu$  in the lab frame can be written as
\begin{equation} 
	\frac{dN(E_\nu,E_N)}{dE_\nu}= 
	\begin{cases} \displaystyle \frac{A_1}{E_N}+\frac{A_2E_\nu}{E_N^2}, & 0<E_\nu<E_N\left(1-\frac{m_V^2}{M_N^2}\right), \\[0.3cm] 0, & \text{otherwise}. \end{cases} 
\end{equation} in which
$$A_1=\frac{m_N(1-\alpha)}{2 \tilde{E}_\nu}  \ \ {\rm and} \ \  A_2=\frac{\alpha m_N^2}{2(\tilde{E}_\nu)^2} \ \ {\rm and} \ \ \tilde{E}_\nu=\frac{m_N^2-m_V^2}{2m_N}$$
with $$
\alpha=\frac{m_N^2-2m_V^2}{m_N^2+2m_V^2}  \ \ {\rm and} \ \ \gamma=
\frac{E_N}{m_N} \ . $$
Moreover the spectrum of $V$ is  
\begin{equation} 
	\frac{dN(E_V,E_N)}{dE_V}= 
	\begin{cases} \displaystyle \frac{m_N}{2 E_N\tilde{E}_\nu}\left[ 1 -\alpha \left( \frac{1-E_V/E_N}{\tilde{E}_\nu/m_N}-1\right)\right], & E_N\frac{m_V^2}{m_N^2}<E_V<E_N, \\[0.3cm] 0, & \text{otherwise}. \end{cases} 
\end{equation} 
Notice that 
$$ \int_{E_\nu }^ {\infty}E_N^{-\gamma_\nu} \frac{dN(E_\nu,E_N)}{dE_\nu} dE_N= \int_{E_\nu \frac{m_N^2}{m_N^2-m_V^2}}^ {\infty}E_N^{-\gamma_\nu} \frac{dN(E_\nu,E_N)}{dE_\nu} dE_N = \mathcal{B} E_\nu^{-\gamma_\nu} \ ,$$
 in which \begin{equation}\mathcal{B}=\left( \frac{m_N^2-m_V^2}{m_N^2}\right)^{\gamma_\nu} \left(\frac{A_1}{\gamma_\nu}+\frac{A_2}{1+\gamma_\nu}  \frac{m_N^2-m_V^2}{m_N^2}\right) .\label{matB}\end{equation}
The energy spectrum of the secondary $\nu$ and $\bar{\nu}$ from $N\to V\nu$ and $V\to \bar{\nu}\nu$ will be of form
\begin{equation}
	\frac{dN(E_\nu,E_N)}{d E_\nu}=\frac{m_N^2}{m_N^2-m_V^2}\frac{1}{E_N}\log \frac{E_N}{{\rm Max}[E_\nu, E_N m_V^2/m_N^2]}
\end{equation}
with which
$$\int_{E_\nu }^ {\infty}E_N^{-\gamma_\nu} \frac{dN(E_\nu,E_N)}{dE_\nu} dE_N =
\frac{m_N^2}{m_N^2-m_V^2} \frac{1-(m_V^2/m_N^2)^{\gamma_\nu}}{\gamma_\nu^2} E_\nu^{-\gamma_\nu}.$$
Let us now consider the two flavor case with $N_e$ and $N_\tau$ respectively mixed with $\nu_e$ and $\nu_\tau$. As discussed  before, we can take the energy of the upscattered $N_e$ or $N_\tau$ equal to the energy of the initial neutrino. Moreover, as discussed in sect. \ref{medium}, for the region in the spike where the density is high enough for efficient scattering, the induced mass for the neutrinos will prevent decoherence so we do not need to worry about the oscillation among the active neutrino flavors.  We can  generalize Eqs. (\ref{Phi-evolution},\ref{initials},\ref{maintain},\ref{b}) for  two flavors and using the above relations, we then arrive at the following differential  equations  for the evolution of the neutrino fluxes: 
 
	\begin{eqnarray} \label{Flux-evolution}
		\frac{d\Phi_{\nu_e}}{d\tau}&=-a\Phi_{\nu_e}+b_{ee}\Phi_{\nu_e} +\bar{b}_{ee}\Phi_{\bar{\nu}_e} +b_{\tau e} (\Phi_{\nu_\tau}+\Phi_{\bar{\nu}_\tau})    \nonumber \\ 
		\frac{d\Phi_{\nu_\tau}}{d\tau}&=-\Phi_{\nu_\tau}+b_{\tau \tau} \Phi_{\nu_\tau}+\bar{b}_{\tau \tau}\Phi_{\bar{\nu}_\tau}+b_{e\tau}(\Phi_{\nu_e}+ \Phi_{\bar{\nu}_e})
	\end{eqnarray}
	in which 
	$$d\tau=\sigma (\nu_\tau \psi \to N_\tau \psi)\frac{\rho_{spike}}{m_\psi}dr  \ \ {\rm and}  \  \ a= \frac{\sigma(\nu_e+\psi\to N_e+\psi)}{\sigma(\nu_\tau+\psi\to N_\tau+\psi)}\simeq \frac{|U_{e 4}|^2}{|U_{\tau 5}|^2}\ ,$$
	$$b_{ee}=a\mathcal{B}_e+\frac{a}{\gamma_\nu^2}Br(V\to \nu_e \bar{\nu}_e) \left(1- \left(\frac{m_V^2}{m_{N_e}^2} \right)^{\gamma_\nu} \right)\frac{m_{N_e}^2}{m_{N_e}^2-m_V^2}\ , $$
	$$\bar{b}_{ee}=
	\frac{a}{\gamma_\nu^2}Br(V\to \nu_e \bar{\nu}_e) \left(1- \left(\frac{m_V^2}{m_{N_e}^2} \right)^{\gamma_\nu} \right)\frac{m_{N_e}^2}{m_{N_e}^2-m_V^2}$$
	$$ b_{\tau \tau}=\mathcal{B}_\tau+
	\frac{1}{\gamma_\nu^2}Br(V\to \nu_\tau \bar{\nu}_\tau)\left(1- \left(\frac{m_V^2}{m_{N_\tau}^2} \right)^{\gamma_\nu} \right)\frac{m_{N_\tau}^2}{m_{N_\tau}^2-m_V^2}\ ,$$
	$$ \bar{b}_{\tau \tau}=	\frac{1}{\gamma_\nu^2}Br(V\to \nu_\tau \bar{\nu}_\tau)\left(1- \left(\frac{m_V^2}{m_{N_\tau}^2} \right)^{\gamma_\nu} \right)\frac{m_{N_\tau}^2}{m_{N_\tau}^2-m_V^2}\
	$$
	
	in which $\mathcal{B}_e$ and  $\mathcal{B}_\tau$ are given by Eq. (\ref{matB}) replacing $N \to N_e$ and $N \to N_\tau$, respectively. Moreover,
	$$b_{\tau e}=\frac{1}{\gamma_\nu^2}Br(V\to \nu_e \bar{\nu}_e) \left(1- \left(\frac{m_V^2}{m_{N_\tau}^2} \right)^{\gamma_\nu} \right)\frac{m_{N_\tau}^2}{m_{N_\tau}^2-m_V^2}\ ,
	$$
	and
	$$b_{e \tau }=\frac{a}{\gamma_\nu^2}Br(V\to \nu_\tau \bar{\nu}_\tau)\left(1- \left(\frac{m_V^2}{m_{N_e}^2} \right)^{\gamma_\nu} \right)\frac{m_{N_e}^2}{m_{N_e}^2-m_V^2}\ .
	$$  The evolution of the fluxes of antineutrinos are given by Eqs. (\ref{Flux-evolution}), replacing $\Phi_\nu$ with $\Phi_{\bar{\nu}}$ .
In our model,
$${\rm Br}(V\to \nu_\tau \bar{\nu}_\tau)=\frac{|U_{\tau 5}|^4}{|U_{e 4}|^4+ |U_{\tau 5}|^4} \ \  \ {\rm and} \ \ \  {\rm Br}(V\to \nu_e \bar{\nu}_e)=\frac{|U_{e 4}|^4}{|U_{e 4}|^4+ |U_{\tau 5}|^4} \ .$$
Similar consideration and formulas hold valid for the 
two flavor case with $N_\mu$ and $N_\tau$  respectively mixed with $\nu_\mu$ and $\nu_\tau$.

\section*{Acknowledgments} 

A M acknowledges the financial support from the Iran National Science Foundation (INSF) through grant 40400765. We would like also to thank S. Ansarifard  for his useful discussion. 
This project has received funding from the European Union's Horizon Europe research and innovation programme under the Marie Sklodowska-Curie Staff Exchange grant agreement No 101086085-ASYMMETRY.
The authors acknowledge the computational resources provided by the Scalable Analytics and Research Virtual Environment (SARVe) facility at the School of Theoretical Physics, IPM.

\bibliographystyle{JHEP}
\bibliography{bibbib}

\providecommand{\href}[2]{#2}\begingroup\raggedright\begin{thebibliography}{10}

\bibitem{Gondolo:1999ef}
P.~Gondolo and J.~Silk, \emph{{Dark matter annihilation at the galactic
  center}}, \href{https://doi.org/10.1103/PhysRevLett.83.1719}{\emph{Phys. Rev.
  Lett.} {\bfseries 83} (1999) 1719}
  [\href{https://arxiv.org/abs/astro-ph/9906391}{{\ttfamily
  astro-ph/9906391}}].

\bibitem{Sharma:2025ynw}
M.~Sharma, G.~Herrera, N.~Arav and S.~Horiuchi, \emph{{Novel method to trace
  the dark matter density profile around supermassive black holes with AGN
  reverberation mapping}}, \href{https://doi.org/10.1103/llpr-gnmh}{\emph{Phys.
  Rev. D} {\bfseries 113} (2026) 043052}
  [\href{https://arxiv.org/abs/2506.10122}{{\ttfamily 2506.10122}}].

\bibitem{Vasiliev:2007vh}
E.~Vasiliev, \emph{{Dark matter annihilation near a black hole: Plateau vs.
  weak cusp}}, \href{https://doi.org/10.1103/PhysRevD.76.103532}{\emph{Phys.
  Rev. D} {\bfseries 76} (2007) 103532}
  [\href{https://arxiv.org/abs/0707.3334}{{\ttfamily 0707.3334}}].

\bibitem{Shapiro:2016ypb}
S.L.~Shapiro and J.~Shelton, \emph{{Weak annihilation cusp inside the dark
  matter spike about a black hole}},
  \href{https://doi.org/10.1103/PhysRevD.93.123510}{\emph{Phys. Rev. D}
  {\bfseries 93} (2016) 123510}
  [\href{https://arxiv.org/abs/1606.01248}{{\ttfamily 1606.01248}}].

\bibitem{Chattopadhyay:2026kbm}
D.S.~Chattopadhyay, P.S.B.~Dev and Y.~Porto, \emph{{Ruling Out Spiky WIMP Dark
  Matter using Indirect Searches}},
  \href{https://arxiv.org/abs/2602.23348}{{\ttfamily 2602.23348}}.

\bibitem{DeMarchi:2025xag}
A.G.~De~Marchi, A.~Granelli, J.~Nava and F.~Sala, \emph{{Diffuse astrophysical
  neutrinos from dark matter around blazars}},
  \href{https://doi.org/10.1016/j.physletb.2025.140015}{\emph{Phys. Lett. B}
  {\bfseries 871} (2025) 140015}
  [\href{https://arxiv.org/abs/2506.06416}{{\ttfamily 2506.06416}}].

\bibitem{DeMarchi:2025uoo}
A.G.~De~Marchi, A.~Granelli, J.~Nava and F.~Sala, \emph{{Boosted dark matter
  versus dark matter-induced neutrinos from single and stacked blazars}},
  \href{https://doi.org/10.1007/JHEP12(2025)136}{\emph{JHEP} {\bfseries 12}
  (2025) 136} [\href{https://arxiv.org/abs/2507.12278}{{\ttfamily
  2507.12278}}].

\bibitem{Fujiwara:2024qos}
M.~Fujiwara, G.~Herrera and S.~Horiuchi, \emph{{Neutrino Diffusion within Dark
  Matter Spikes}},  \href{https://arxiv.org/abs/2412.00805}{{\ttfamily
  2412.00805}}.

\bibitem{Cline:2022qld}
J.M.~Cline, S.~Gao, F.~Guo, Z.~Lin, S.~Liu, M.~Puel et~al., \emph{{Blazar
  Constraints on Neutrino-Dark Matter Scattering}},
  \href{https://doi.org/10.1103/PhysRevLett.130.091402}{\emph{Phys. Rev. Lett.}
  {\bfseries 130} (2023) 091402}
  [\href{https://arxiv.org/abs/2209.02713}{{\ttfamily 2209.02713}}].

\bibitem{Cline:2023tkp}
J.M.~Cline and M.~Puel, \emph{{NGC 1068 constraints on neutrino-dark matter
  scattering}},
  \href{https://doi.org/10.1088/1475-7516/2023/06/004}{\emph{JCAP} {\bfseries
  06} (2023) 004} [\href{https://arxiv.org/abs/2301.08756}{{\ttfamily
  2301.08756}}].

\bibitem{Zapata:2025huq}
G.D.~Zapata, J.~Jones-P{\'e}rez and A.M.~Gago, \emph{{Bounds on neutrino-DM
  interactions from TXS~0506+056 neutrino outburst}},
  \href{https://doi.org/10.1088/1475-7516/2025/07/042}{\emph{JCAP} {\bfseries
  07} (2025) 042} [\href{https://arxiv.org/abs/2503.03823}{{\ttfamily
  2503.03823}}].

\bibitem{Dixit:2026zkv}
K.~Dixit, G.~Mohlabeng and S.~Razzaque, \emph{{Searching for dark matter
  signals with high energy astrophysical neutrinos in IceCube}},
  \href{https://arxiv.org/abs/2602.06121}{{\ttfamily 2602.06121}}.

\bibitem{Mondol:2025uuw}
R.~Mondol, S.~Bouri, A.K.~Saha and R.~Laha, \emph{{Road through Dark$\nu$ess:
  Probing dark matter-neutrino interactions using KM3-230213A}},
  \href{https://arxiv.org/abs/2506.19910}{{\ttfamily 2506.19910}}.

\bibitem{Nussinov:1985xr}
S.~Nussinov, \emph{{TECHNOCOSMOLOGY: COULD A TECHNIBARYON EXCESS PROVIDE A
  'NATURAL' MISSING MASS CANDIDATE?}},
  \href{https://doi.org/10.1016/0370-2693(85)90689-6}{\emph{Phys. Lett. B}
  {\bfseries 165} (1985) 55}.

\bibitem{Barr:1990ca}
S.M.~Barr, R.S.~Chivukula and E.~Farhi, \emph{{Electroweak Fermion Number
  Violation and the Production of Stable Particles in the Early Universe}},
  \href{https://doi.org/10.1016/0370-2693(90)91661-T}{\emph{Phys. Lett. B}
  {\bfseries 241} (1990) 387}.

\bibitem{Barr:1991qn}
S.M.~Barr, \emph{{Baryogenesis, sphalerons and the cogeneration of dark
  matter}}, \href{https://doi.org/10.1103/PhysRevD.44.3062}{\emph{Phys. Rev. D}
  {\bfseries 44} (1991) 3062}.

\bibitem{Kaplan:1991ah}
D.B.~Kaplan, \emph{{A Single explanation for both the baryon and dark matter
  densities}}, \href{https://doi.org/10.1103/PhysRevLett.68.741}{\emph{Phys.
  Rev. Lett.} {\bfseries 68} (1992) 741}.

\bibitem{Davoudiasl:2012uw}
H.~Davoudiasl and R.N.~Mohapatra, \emph{{On Relating the Genesis of Cosmic
  Baryons and Dark Matter}},
  \href{https://doi.org/10.1088/1367-2630/14/9/095011}{\emph{New J. Phys.}
  {\bfseries 14} (2012) 095011}
  [\href{https://arxiv.org/abs/1203.1247}{{\ttfamily 1203.1247}}].

\bibitem{Zurek:2013wia}
K.M.~Zurek, \emph{{Asymmetric Dark Matter: Theories, Signatures, and
  Constraints}},
  \href{https://doi.org/10.1016/j.physrep.2013.12.001}{\emph{Phys. Rept.}
  {\bfseries 537} (2014) 91} [\href{https://arxiv.org/abs/1308.0338}{{\ttfamily
  1308.0338}}].

\bibitem{Dev:2026ddq}
P.S.B.~Dev, J.~Gehrlein, A.~Sengupta and A.~Soni, \emph{{A Minimal Dark $SU(2)$
  Origin of a Massless Dirac Neutrino}},
  \href{https://arxiv.org/abs/2605.28923}{{\ttfamily 2605.28923}}.

\bibitem{Inoue:2019yfs}
Y.~Inoue, D.~Khangulyan and A.~Doi, \emph{{On the Origin of High-energy
  Neutrinos from NGC 1068: The Role of Nonthermal Coronal Activity}},
  \href{https://doi.org/10.3847/2041-8213/ab7661}{\emph{Astrophys. J. Lett.}
  {\bfseries 891} (2020) L33}
  [\href{https://arxiv.org/abs/1909.02239}{{\ttfamily 1909.02239}}].

\bibitem{Murase:2019vdl}
K.~Murase, S.S.~Kimura and P.~Meszaros, \emph{{Hidden Cores of Active Galactic
  Nuclei as the Origin of Medium-Energy Neutrinos: Critical Tests with the MeV
  Gamma-Ray Connection}},
  \href{https://doi.org/10.1103/PhysRevLett.125.011101}{\emph{Phys. Rev. Lett.}
  {\bfseries 125} (2020) 011101}
  [\href{https://arxiv.org/abs/1904.04226}{{\ttfamily 1904.04226}}].

\bibitem{Kheirandish:2021wkm}
A.~Kheirandish, K.~Murase and S.S.~Kimura, \emph{{High-energy Neutrinos from
  Magnetized Coronae of Active Galactic Nuclei and Prospects for Identification
  of Seyfert Galaxies and Quasars in Neutrino Telescopes}},
  \href{https://doi.org/10.3847/1538-4357/ac1c77}{\emph{Astrophys. J.}
  {\bfseries 922} (2021) 45}
  [\href{https://arxiv.org/abs/2102.04475}{{\ttfamily 2102.04475}}].

\bibitem{Murase:2022dog}
K.~Murase, \emph{{Hidden Hearts of Neutrino Active Galaxies}},
  \href{https://doi.org/10.3847/2041-8213/aca53c}{\emph{Astrophys. J. Lett.}
  {\bfseries 941} (2022) L17}
  [\href{https://arxiv.org/abs/2211.04460}{{\ttfamily 2211.04460}}].

\bibitem{Eichmann:2022lxh}
B.~Eichmann, F.~Oikonomou, S.~Salvatore, R.-J.~Dettmar and J.~Becker~Tjus,
  \emph{{Solving the Multimessenger Puzzle of the AGN-starburst Composite
  Galaxy NGC 1068}},
  \href{https://doi.org/10.3847/1538-4357/ac9588}{\emph{Astrophys. J.}
  {\bfseries 939} (2022) 43}
  [\href{https://arxiv.org/abs/2207.00102}{{\ttfamily 2207.00102}}].

\bibitem{Inoue:2022yak}
S.~Inoue, M.~Cerruti, K.~Murase and R.-Y.~Liu, \emph{{Multimessenger emission
  from winds and tori in active galactic nuclei}},
  \href{https://doi.org/10.22323/1.444.1161}{\emph{PoS} {\bfseries ICRC2023}
  (2023) 1161} [\href{https://arxiv.org/abs/2207.02097}{{\ttfamily
  2207.02097}}].

\bibitem{Yoast-Hull:2013qfa}
T.M.~Yoast-Hull, J.S.G.~III, E.G.~Zweibel and J.E.~Everett, \emph{{Active
  Galactic Nuclei, Neutrinos, and Interacting Cosmic Rays in NGC 253 and NGC
  1068}}, \href{https://doi.org/10.1088/0004-637X/780/2/137}{\emph{Astrophys.
  J.} {\bfseries 780} (2014) 137}
  [\href{https://arxiv.org/abs/1311.5586}{{\ttfamily 1311.5586}}].

\bibitem{Fiorillo:2023dts}
D.F.G.~Fiorillo, M.~Petropoulou, L.~Comisso, E.~Peretti and L.~Sironi,
  \emph{{TeV Neutrinos and Hard X-Rays from Relativistic Reconnection in the
  Corona of NGC 1068}},
  \href{https://doi.org/10.3847/2041-8213/ad192b}{\emph{Astrophys. J.}
  {\bfseries 961} (2024) L14}
  [\href{https://arxiv.org/abs/2310.18254}{{\ttfamily 2310.18254}}].

\bibitem{Fiorillo:2024akm}
D.F.G.~Fiorillo, L.~Comisso, E.~Peretti, M.~Petropoulou and L.~Sironi, \emph{{A
  Magnetized Strongly Turbulent Corona as the Source of Neutrinos from NGC
  1068}}, \href{https://doi.org/10.3847/1538-4357/ad7021}{\emph{Astrophys. J.}
  {\bfseries 974} (2024) 75}
  [\href{https://arxiv.org/abs/2407.01678}{{\ttfamily 2407.01678}}].

\bibitem{Ullio:2001fb}
P.~Ullio, H.~Zhao and M.~Kamionkowski, \emph{{A Dark matter spike at the
  galactic center?}},
  \href{https://doi.org/10.1103/PhysRevD.64.043504}{\emph{Phys. Rev. D}
  {\bfseries 64} (2001) 043504}
  [\href{https://arxiv.org/abs/astro-ph/0101481}{{\ttfamily
  astro-ph/0101481}}].

\bibitem{Balaji:2023hmy}
S.~Balaji, D.~Sachdeva, F.~Sala and J.~Silk, \emph{{Dark matter spikes around
  Sgr A* in {\ensuremath{\gamma}}-rays}},
  \href{https://doi.org/10.1088/1475-7516/2023/08/063}{\emph{JCAP} {\bfseries
  08} (2023) 063} [\href{https://arxiv.org/abs/2303.12107}{{\ttfamily
  2303.12107}}].

\bibitem{Gnedin:2003rj}
O.Y.~Gnedin and J.R.~Primack, \emph{{Dark Matter Profile in the Galactic
  Center}}, \href{https://doi.org/10.1103/PhysRevLett.93.061302}{\emph{Phys.
  Rev. Lett.} {\bfseries 93} (2004) 061302}
  [\href{https://arxiv.org/abs/astro-ph/0308385}{{\ttfamily
  astro-ph/0308385}}].

\bibitem{Shapiro:2022prq}
S.L.~Shapiro and D.C.~Heggie, \emph{{Effect of stars on the dark matter spike
  around a black hole: A tale of two treatments}},
  \href{https://doi.org/10.1103/PhysRevD.106.043018}{\emph{Phys. Rev. D}
  {\bfseries 106} (2022) 043018}
  [\href{https://arxiv.org/abs/2209.08105}{{\ttfamily 2209.08105}}].

\bibitem{Bertone:2005hw}
G.~Bertone and D.~Merritt, \emph{{Time-dependent models for dark matter at the
  Galactic Center}},
  \href{https://doi.org/10.1103/PhysRevD.72.103502}{\emph{Phys. Rev. D}
  {\bfseries 72} (2005) 103502}
  [\href{https://arxiv.org/abs/astro-ph/0501555}{{\ttfamily
  astro-ph/0501555}}].

\bibitem{IceCube:2022der}
{\scshape IceCube} collaboration, \emph{{Evidence for neutrino emission from
  the nearby active galaxy NGC 1068}},
  \href{https://doi.org/10.1126/science.abg3395}{\emph{Science} {\bfseries 378}
  (2022) 538} [\href{https://arxiv.org/abs/2211.09972}{{\ttfamily
  2211.09972}}].

\bibitem{IceCube:2024ayt}
{\scshape IceCube} collaboration, \emph{{Search for Neutrino Emission from Hard
  X-Ray AGN with IceCube}},
  \href{https://doi.org/10.3847/1538-4357/ada94b}{\emph{Astrophys. J.}
  {\bfseries 981} (2025) 131}
  [\href{https://arxiv.org/abs/2406.06684}{{\ttfamily 2406.06684}}].

\bibitem{IceCube:2018dnn}
{\scshape IceCube, Fermi-LAT, MAGIC, AGILE, ASAS-SN, HAWC, H.E.S.S., INTEGRAL,
  Kanata, Kiso, Kapteyn, Liverpool Telescope, Subaru, Swift NuSTAR, VERITAS,
  VLA/17B-403} collaboration, \emph{{Multimessenger observations of a flaring
  blazar coincident with high-energy neutrino IceCube-170922A}},
  \href{https://doi.org/10.1126/science.aat1378}{\emph{Science} {\bfseries 361}
  (2018) eaat1378} [\href{https://arxiv.org/abs/1807.08816}{{\ttfamily
  1807.08816}}].

\bibitem{Greenhill:1996te}
L.J.~Greenhill, C.R.~Gwinn, R.~Antonucci and R.~Barvainis, \emph{{Vlbi imaging
  of water maser emission from the nuclear torus of ngc 1068}},
  \href{https://doi.org/10.1086/310346}{\emph{Astrophys. J. Lett.} {\bfseries
  472} (1996) L21} [\href{https://arxiv.org/abs/astro-ph/9609082}{{\ttfamily
  astro-ph/9609082}}].

\bibitem{Bentz:2022NGC4151}
M.C.~Bentz, P.R.~Williams and T.~Treu, \emph{The broad line region and black
  hole mass of ngc 4151},
  \href{https://doi.org/10.3847/1538-4357/ac7c74}{\emph{The Astrophysical
  Journal} {\bfseries 934} (2022) 168}.

\bibitem{Padovani:2019xcv}
P.~Padovani, F.~Oikonomou, M.~Petropoulou, P.~Giommi and E.~Resconi, \emph{{TXS
  0506+056, the first cosmic neutrino source, is not a BL Lac}},
  \href{https://doi.org/10.1093/mnrasl/slz011}{\emph{Mon. Not. Roy. Astron.
  Soc.} {\bfseries 484} (2019) L104}
  [\href{https://arxiv.org/abs/1901.06998}{{\ttfamily 1901.06998}}].

\bibitem{Cerruti:2017mnz}
M.~Cerruti, W.~Benbow, X.~Chen, J.P.~Dumm, L.F.~Fortson and K.~Shahinyan,
  \emph{{Luminous and high-frequency peaked blazars: the origin of the
  {\ensuremath{\gamma}}-ray emission from PKS 1424+240}},
  \href{https://doi.org/10.1051/0004-6361/201730799}{\emph{Astron. Astrophys.}
  {\bfseries 606} (2017) A68}
  [\href{https://arxiv.org/abs/1707.00804}{{\ttfamily 1707.00804}}].

\bibitem{Dev:2025tdv}
P.S.B.~Dev, D.~Kim, D.~Sathyan, K.~Sinha and Y.~Zhang, \emph{{New Constraints
  on Neutrino-Dark Matter Interactions: A Comprehensive Analysis}},
  \href{https://arxiv.org/abs/2507.01000}{{\ttfamily 2507.01000}}.

\bibitem{Farzan:2026wek}
Y.~Farzan, \emph{{Revisiting bounds on neutrino dark matter interaction at
  spikes}},  \href{https://arxiv.org/abs/2607.08904}{{\ttfamily 2607.08904}}.

\bibitem{Pushkarev:2017fbk}
A.B.~Pushkarev, Y.Y.~Kovalev, M.L.~Lister and T.~Savolainen, \emph{{MOJAVE
  {\textendash} XIV. Shapes and opening angles of AGN jets}},
  \href{https://doi.org/10.1093/mnras/stx854}{\emph{Mon. Not. Roy. Astron.
  Soc.} {\bfseries 468} (2017) 4992}
  [\href{https://arxiv.org/abs/1705.02888}{{\ttfamily 1705.02888}}].

\bibitem{Barillier:2025xct}
E.~Barillier, L.~Manenti, K.~Mora, P.~Padovani, I.~Sarnoff, Y.~Xu et~al.,
  \emph{{Setting limits on blazar-boosted dark matter with xenon-based
  detectors}}, \href{https://doi.org/10.1103/qxkb-4bpy}{\emph{Phys. Rev. D}
  {\bfseries 113} (2026) 023005}
  [\href{https://arxiv.org/abs/2509.07265}{{\ttfamily 2509.07265}}].

\bibitem{NOMAD:2001eyx}
{\scshape NOMAD} collaboration, \emph{{Search for heavy neutrinos mixing with
  tau neutrinos}},
  \href{https://doi.org/10.1016/S0370-2693(01)00362-8}{\emph{Phys. Lett. B}
  {\bfseries 506} (2001) 27}
  [\href{https://arxiv.org/abs/hep-ex/0101041}{{\ttfamily hep-ex/0101041}}].

\bibitem{Orloff:2002de}
J.~Orloff, A.N.~Rozanov and C.~Santoni, \emph{{Limits on the mixing of tau
  neutrino to heavy neutrinos}},
  \href{https://doi.org/10.1016/S0370-2693(02)02769-7}{\emph{Phys. Lett. B}
  {\bfseries 550} (2002) 8}
  [\href{https://arxiv.org/abs/hep-ph/0208075}{{\ttfamily hep-ph/0208075}}].

\bibitem{DELPHI:1996qcc}
{\scshape DELPHI} collaboration, \emph{{Search for neutral heavy leptons
  produced in Z decays}}, \href{https://doi.org/10.1007/s002880050370}{\emph{Z.
  Phys. C} {\bfseries 74} (1997) 57}.

\bibitem{ArgoNeuT:2021clc}
{\scshape ArgoNeuT} collaboration, \emph{{New Constraints on Tau-Coupled Heavy
  Neutral Leptons with Masses mN=280{\textendash}970{\,}{\,}MeV}},
  \href{https://doi.org/10.1103/PhysRevLett.127.121801}{\emph{Phys. Rev. Lett.}
  {\bfseries 127} (2021) 121801}
  [\href{https://arxiv.org/abs/2106.13684}{{\ttfamily 2106.13684}}].

\bibitem{BaBar:2022cqj}
{\scshape BaBar} collaboration, \emph{{Search for heavy neutral leptons using
  tau lepton decays at BaBaR}},
  \href{https://doi.org/10.1103/PhysRevD.107.052009}{\emph{Phys. Rev. D}
  {\bfseries 107} (2023) 052009}
  [\href{https://arxiv.org/abs/2207.09575}{{\ttfamily 2207.09575}}].

\bibitem{Bilenky:1992xn}
M.S.~Bilenky, S.M.~Bilenky and A.~Santamaria, \emph{{Invisible width of the Z
  boson and 'secret' neutrino-neutrino interactions}},
  \href{https://doi.org/10.1016/0370-2693(93)90703-K}{\emph{Phys. Lett. B}
  {\bfseries 301} (1993) 287}.

\bibitem{Berryman:2022hds}
J.M.~Berryman et~al., \emph{{Neutrino self-interactions: A white paper}},
  \href{https://doi.org/10.1016/j.dark.2023.101267}{\emph{Phys. Dark Univ.}
  {\bfseries 42} (2023) 101267}
  [\href{https://arxiv.org/abs/2203.01955}{{\ttfamily 2203.01955}}].

\bibitem{Bakhti:2016prn}
P.~Bakhti and Y.~Farzan, \emph{{CP-Violation and Non-Standard Interactions at
  the MOMENT}}, \href{https://doi.org/10.1007/JHEP07(2016)109}{\emph{JHEP}
  {\bfseries 07} (2016) 109}
  [\href{https://arxiv.org/abs/1602.07099}{{\ttfamily 1602.07099}}].

\bibitem{Fernandez-Martinez:2016lgt}
E.~Fernandez-Martinez, J.~Hernandez-Garcia and J.~Lopez-Pavon, \emph{{Global
  constraints on heavy neutrino mixing}},
  \href{https://doi.org/10.1007/JHEP08(2016)033}{\emph{JHEP} {\bfseries 08}
  (2016) 033} [\href{https://arxiv.org/abs/1605.08774}{{\ttfamily
  1605.08774}}].

\bibitem{NA62:2020mcv}
{\scshape NA62} collaboration, \emph{{Search for heavy neutral lepton
  production in K+ decays to positrons}},
  \href{https://doi.org/10.1016/j.physletb.2020.135599}{\emph{Phys. Lett. B}
  {\bfseries 807} (2020) 135599}
  [\href{https://arxiv.org/abs/2005.09575}{{\ttfamily 2005.09575}}].

\bibitem{Liu:2025akp}
Y.-B.~Liu and J.-W.~Lian, \emph{{Searches for heavy neutrinos at a 3 TeV CLIC
  in fat jet final states}},
  \href{https://doi.org/10.1088/1674-1137/ae2d24}{\emph{Chin. Phys. C}
  {\bfseries 50} (2026) 033110}
  [\href{https://arxiv.org/abs/2511.02557}{{\ttfamily 2511.02557}}].

\bibitem{Chauhan:2025hoz}
G.~Chauhan, R.A.~Gustafson, G.~Herrera, T.~Johnson and I.M.~Shoemaker,
  \emph{{The dark matter diffused supernova neutrino background}},
  \href{https://doi.org/10.1088/1475-7516/2025/10/020}{\emph{JCAP} {\bfseries
  10} (2025) 020} [\href{https://arxiv.org/abs/2505.03882}{{\ttfamily
  2505.03882}}].

\bibitem{Esteban:2025wbv}
I.~Esteban and A.~Ibarra, \emph{{Attenuation of the ultra-high-energy neutrino
  flux by dark matter scatterings}},
  \href{https://doi.org/10.1088/1475-7516/2026/04/064}{\emph{JCAP} {\bfseries
  04} (2026) 064} [\href{https://arxiv.org/abs/2508.02869}{{\ttfamily
  2508.02869}}].

\bibitem{Bertolez-Martinez:2025trs}
T.~Bert{\'o}lez-Mart{\'\i}nez, G.~Herrera, P.~Mart{\'\i}nez-Mirav{\'e} and
  J.~Terol~Calvo, \emph{{The Highest-Energy Neutrino Event Constrains Dark
  Matter-Neutrino Interactions}},
  \href{https://arxiv.org/abs/2506.08993}{{\ttfamily 2506.08993}}.

\bibitem{Abbasi:2025fjc}
R.~Abbasi et~al., \emph{{Characterization of the Three-Flavor Composition of
  Cosmic Neutrinos with IceCube}},
  \href{https://arxiv.org/abs/2510.24957}{{\ttfamily 2510.24957}}.

\bibitem{IceCube-Gen2:2020qha}
{\scshape IceCube-Gen2} collaboration, \emph{{IceCube-Gen2: the window to the
  extreme Universe}}, \href{https://doi.org/10.1088/1361-6471/abbd48}{\emph{J.
  Phys. G} {\bfseries 48} (2021) 060501}
  [\href{https://arxiv.org/abs/2008.04323}{{\ttfamily 2008.04323}}].

\bibitem{IceCube-Gen2:2025yte}
{\scshape IceCube-Gen2} collaboration, \emph{{Probing ultra-high-energy
  neutrinos with the IceCube-Gen2 in-ice radio array}},
  \href{https://doi.org/10.22323/1.501.1045}{\emph{PoS} {\bfseries ICRC2025}
  (2025) 1045} [\href{https://arxiv.org/abs/2507.07813}{{\ttfamily
  2507.07813}}].

\bibitem{Wen:2026ngx}
A.Y.~Wen, C.A.~Arg{\"u}elles and S.~Palomares-Ruiz, \emph{{Visible inelasticity
  as a probe of tau flavor content of astrophysical neutrinos}},
  \href{https://arxiv.org/abs/2605.29105}{{\ttfamily 2605.29105}}.

\bibitem{Skrzypek:2025tmg}
B.~Skrzypek and S.~Klein, \emph{{Measuring the Astrophysical
  Neutrino-Antineutrino Ratio with IceCube}},
  \href{https://doi.org/10.22323/1.501.1181}{\emph{PoS} {\bfseries ICRC2025}
  (2025) 1181} [\href{https://arxiv.org/abs/2508.02795}{{\ttfamily
  2508.02795}}].

\bibitem{Kim-Giunti}
C.~Giunti and C.W.~Kim, \emph{Fundamentals of neutrino physics and
  astrophysics}, Oxford Univ Press, New York (2011).

\bibitem{Farzan:2018pnk}
Y.~Farzan and S.~Palomares-Ruiz, \emph{{Flavor of cosmic neutrinos preserved by
  ultralight dark matter}},
  \href{https://doi.org/10.1103/PhysRevD.99.051702}{\emph{Phys. Rev. D}
  {\bfseries 99} (2019) 051702}
  [\href{https://arxiv.org/abs/1810.00892}{{\ttfamily 1810.00892}}].

\bibitem{Hajjar:2026uzs}
R.~Hajjar, M.~Bustamante and S.~Palomares-Ruiz, \emph{{Measurement of the
  neutrino cross section across the TeV scale using IceCube data}},
  \href{https://arxiv.org/abs/2610.24068}{{\ttfamily 2610.24068}}.

\bibitem{Herrera:2026hal}
G.~Herrera, A.~Hussein, L.~Necib, E.Y.~Davies and X.~Shen, \emph{{Formation and
  Redshift Evolution of Dark Matter Spikes}},
  \href{https://arxiv.org/abs/2605.01023}{{\ttfamily 2605.01023}}.

\bibitem{Bertone:2024rxe}
G.~Bertone, \emph{{Dark matter, black holes, and gravitational waves}},
  \href{https://doi.org/10.1016/j.nuclphysb.2024.116487}{\emph{Nucl. Phys. B}
  {\bfseries 1003} (2024) 116487}
  [\href{https://arxiv.org/abs/2404.11513}{{\ttfamily 2404.11513}}].

\end{thebibliography}\endgroup

\end{document}